# A statistical mechanics investigation of Unfolded Protein Response across organisms

Nicole Luchetti[1,2,*], Keith M. Smith[3], Margherita A. G. Matarrese[1], Alessandro Loppini[4], Simonetta Filippi[1,5,6,*], Letizia Chiodo[1]

[1] Department of Engineering, Università Campus Bio-Medico di Roma, Via Alvaro del Portillo 21, 00128 Rome, Italy
[2] Center for Life Nano- & Neuro-Science, Istituto Italiano di Tecnologia, Viale Regina Elena 291, 00161 Rome, Italy
[3] Department of Sciences, University of Strathclyde, 16 Richmond Street, Glasgow, G1 1XQ, Scotland, UK
[4] Department of Medicine and Surgery, Università Campus Bio-Medico di Roma, Via Alvaro del Portillo 21, 00128 Rome, Italy
[5] National Institute of Optics, National Research Council, Largo Enrico Fermi 6, 50125 Florence, Italy
[6] International Center for Relativistic Astrophysics Network, Piazza della Repubblica 10, 65122 Pescara, Italy

*Corresponding authors: n.luchetti@unicampus.it, s.filippi@unicampus.it

## Abstract

Living systems rely on coordinated molecular interactions, especially those related to gene expression and protein activity. The Unfolded Protein Response is a crucial mechanism in eukaryotic cells, activated when unfolded proteins exceed a critical threshold. It maintains cell homeostasis by enhancing protein folding, initiating quality control, and activating degradation pathways when damage is irreversible. This response functions as a dynamic signaling network, with proteins as nodes and their interactions as edges. We analyze these protein-protein networks across different organisms to understand their intricate intra-cellular interactions and behaviors.

In this work, analyzing twelve organisms, we assess how fundamental measures in network theory can individuate seed-proteins and specific pathways across organisms. We employ network robustness to evaluate and compare the strength of the investigated PPI networks, and the structural controllability of complex networks to find and compare the sets of driver nodes necessary to control the overall networks. We find that network measures are related to phylogenetics, and advanced network methods can identify main pathways of significance in the complete Unfolded Protein Response mechanism.

## Introduction

The Unfolded Protein Response (UPR) [1] is a mechanism adopted by cells to maintain homeostasis within the endoplasmic reticulum (ER) compartment in response to an accumulation of unfolded or improperly folded proteins (**Figure 1**) [2–4]. When protein concentration exceeds physiological levels, pro-survival mechanisms are activated to restore the balance between folded and unfolded proteins [5–7]. The heat shock protein family A member 5 (HSPA5), also known as binding immunoglobulin protein (BiP) [8], is a key promotor of the UPR, activating three stress sensors in the ER: the activating transcription factor 6 (ATF6), the endoplasmic reticulum to nucleus signaling 1 (ERN1), and the eukaryotic translation initiation factor 2 alpha kinase 3 (EIF2AK3), respectively [9–11]. If the adaptive UPR response fails, other pathways are activated, leading to apoptosis and autophagy [6,12]. This mechanism is essential for cell survival in mammals [1,13,14] and is strongly preserved across various organisms, from mammals to yeasts and worms [15–17], as well as in fungi [18] and plants [19,20].

The advancement of network theory has significantly contributed to the study of biological networks, particularly protein-protein interaction (PPI) networks [21–24]. Indeed, by conducting network modeling and topological analysis, researchers can gain insights into genes and proteins involved in various biological functions and disease mechanisms [25–27] The UPR pathway can be described as a PPI network [28–30] and analyzed using complex network tools [31–36]. Information of protein interactions are stored in public databases [37–40], obtained via direct and indirect information (i.e., obtained from experimental Y2H test and homology). Classic measures in network theory, whose definition is briefly reported in **Table 1**, provide valuable insights into network structure and function, but they do not adequately address the dynamic aspects of network behavior and vulnerability to disruptions. Therefore, structural controllability [41,42] and network robustness [43–45] can be used to identify driver nodes and exploit whether the network withstands failures or attacks.

In this study, we study the properties of the ER stress response network in twelve different organisms, to determine if network analysis methods can provide insight into characteristics of PPI networks. Specifically, we want to identify and analyze the factors that impact the "strength" of various UPR networks and their resistance to potential alterations. This includes looking at random-attack strategies and various metric-based attack strategies and identifying similarities between different organism models. Our findings indicate that the various methods we adopt can uncover different network characteristics, such as phylogenetic similarities [1], distinguishing mammals and their animal models, and the identification of relevant molecular pathways within the UPR mechanism across organisms. Thus, we hypothesize that these network methods can be applied more widely to characterize unknown PPI networks in silico.

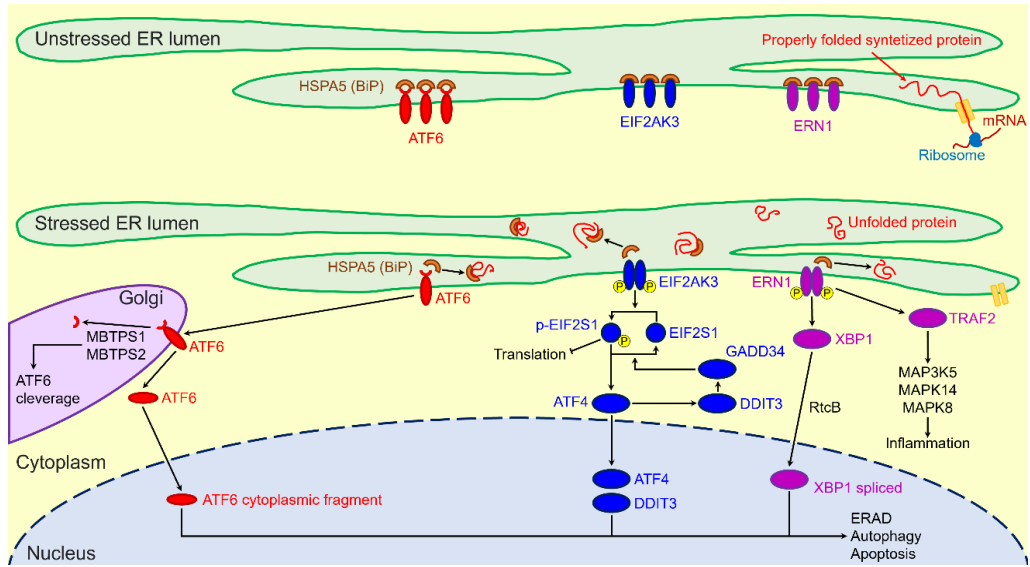

*Figure 1: Visual representation of the main UPR signaling pathways during ER stress in mammals. The cellular process of post-translational modification and protein folding becomes strained, leading to the buildup of non-properly folded proteins. This accumulation can eventually trigger cell death. To counteract ER stress and regain homeostasis, the cell initiates a cascade of signaling pathways. These pathways enhance the production of proteins involved in proper protein folding or facilitate the removal of misfolded proteins through the Endoplasmic-reticulum-associated protein degradation (ERAD).*

*Table 1: Definition of quantities used to describe PPI networks.*

| Term / metric | Definition |
|---|---|
| Barycenter | The node with the lowest value of eccentricity (for us, the absolute center of the networks) |
| Betweenness centrality [46] | Measure of how often a node occur on the shortest paths between other nodes |
| Closeness centrality | $C(x)=(\sum_y d(x,y))^{-1}$<br>is a measure of how close a node is to all other nodes in the network |
| Clustering coefficient | Proportion of edges between the nodes within the $i^{th}$ neighborhood divided by the number of links that could possibly exist between them |
| Average clustering coefficient | $CC = (\sum_i CC_i/n)$<br>is the arithmetic mean of clustering coefficient of all the nodes |
| Density [47] | D=2M/[N(N-1)]<br>where M is the total number of connections in an N nodes network |
| Degree | Number of edges of one node |
| Average Degree | Arithmetic mean of degrees of all network nodes |
| Diameter | It is defined as the eccentricity of a node with the maximum distance to the other nodes |
| Edges | Physical or functional connections between pairs of proteins |
| Modules or Communities | Sub networks that include a high number of inside-sub network edges and a low number of between-sub network edges |
| Modularity | A measure of network tendencies to divide in communities. |
| Nodes | Proteins composing the network |
| Shortest path length | Number of edges needed to connect every pair of nodes through their shortest path |

## Results

The results are presented in three subsections, each elucidating the potential descriptive and predictive power of the methods employed. These methods allow the association of network properties with phylogenetic analogies and assist in identifying biological weaknesses through advanced network descriptors. These subsections correspond to the methods tested in this study: i) standard network descriptors [48], as well as topological analysis [49], ii) robustness [43–45], and iii) structural controllability [41,42]. The methodological pipeline starts by establishing a native UPR model network for each organism, as detailed in **Table 2**, utilizing PPI data sourced from public databases. We then create configuration models by randomizing connections while preserving the same number of connections *per* protein. Well-established network theory measures and advanced network methods are then applied to evaluate the networks for each organism and model.

Table 2: Average values of common network features for all native models.

| | Barycenter | #nodes | #edges | Density | Diameter | Degree | Closeness | Betweenness | Cust. coeff. | Modularity | Communities |
|---|---|---|---|---|---|---|---|---|---|---|---|
| Homo sapiens [2,52] | HSPA5 | 216 | 5286 | 0.114 | 7 | 25 | 0.0020 | 145.5 | 0.522 | 0.311 | 5 |
| Rattus norvegicus [53,54] | Hspa5 | 139 | 2224 | 0.116 | 5 | 16 | 0.0030 | 95.3 | 0.533 | 0.368 | 5 |
| Mus musculus [55–57] | Hspa5 | 172 | 3200 | 0.109 | 6 | 19 | 0.0025 | 119.1 | 0.548 | 0.352 | 6 |
| Macaca fascicularis [58] | HSPA5 | 64 | 578 | 0.143 | 6 | 9 | 0.0067 | 45.8 | 0.544 | 0.419 | 5 |
| Bos taurus [59] | HSPA5 | 93 | 1102 | 0.129 | 5 | 12 | 0.0047 | 64.2 | 0.528 | 0.398 | 4 |
| Oryctologus cuniculis [60,61] | HSPA5 | 54 | 528 | 0.184 | 5 | 10 | 0.0086 | 33.8 | 0.597 | 0.334 | 4 |
| Gallus gallus [62,63] | HSPA5 | 29 | 252 | 0.310 | 4 | 9 | 0.0199 | 12.0 | 0.646 | 0.273 | 2 |
| Danio rerio [64–66] | Hspa5 | 97 | 1058 | 0.114 | 7 | 11 | 0.0042 | 76.9 | 0.560 | 0.371 | 5 |
| c [67–69] | Hsc70-3 | 57 | 760 | 0.238 | 5 | 14 | 0.0089 | 29.9 | 0.560 | 0.324 | 3 |
| Caenorhabditis elegans [70–72] | hsp-90 | 109 | 1212 | 0.103 | 7 | 11 | 0.0034 | 98.9 | 0.594 | 0.450 | 6 |
| Saccharomyces cerevisiae [16,73,74] | KAR2 | 150 | 2834 | 0.127 | 6 | 19 | 0.0028 | 108.0 | 0.607 | 0.352 | 6 |
| Arabidopsis thaliana [75–78] | BIP2; BIP3 | 62 | 1188 | 0.314 | 5 | 19 | 0.0092 | 25.6 | 0.680 | 0.245 | 3 |

### a. Network and topological analyses

*Standard Network Characteristics*

***Barycenter***. We find that the barycenter of all network models corresponds to the Binding Immunoglobulin Protein (KAR2 and BIP2/BIP3 proteins in Saccharomyces cerevisiae and Arabidopsis thaliana are homologous to mammal HSPA5), apart from Caenorhabditis elegans, for which the Heat shock protein 90 (hsp-90) results as the key protein (**Table 2**). Indeed, literature shows that in Caenorhabditis elegans, hsp-90 plays a crucial role in the chemotaxis to non-volatile and volatile attractants detected by AWC sensory neurons [50,51].

***Density***. Arabidopsis thaliana, Gallus gallus and Drosophila melanogaster are the most densely connected networks, with the values of 0.314, 0.310 and 0.238, respectively (**Table 2**). The less densely connected network is found for Caenorhabditis elegans (0.103) and Mus musculus (0.109). Interestingly, we note a threshold value in the number of nodes (about 100 nodes) above which we have a linear trend of both average and normalized average degrees. This may be since increasing the size of the network can provide a decrease in the available public information for some organisms rather than others.

***Average Degree***. The highest average degree value of the native models is observed for Homo sapiens (25) followed by Mus musculus, Arabidopsis thaliana and Saccharomyces cerevisiae with a value of average degree of 19. The lowest average degree value is found for Macaca fascicularis and Gallus gallus, with a numeric value of 9, followed by Oryctologus cuniculis with a value of 10.

***Closeness centrality***. Gallus gallus has the highest closeness, with a value of $19.9 \cdot 10^{-3}$, followed by Arabidopsis thaliana ($9.2 \cdot 10^{-3}$) and Drosophila melanogaster ($8.9 \cdot 10^{-3}$). Homo sapiens shows the smallest value of closeness, with a numeric value of $2.0 \cdot 10^{-3}$.

***Betweenness centrality***. As the average degree, the highest betweenness is observed for Homo sapiens (145.5) followed by Mus musculus (119.1) and Saccharomyces cerevisiae (108.0). On the contrary, Gallus gallus (12.0), Arabidopsis thaliana (25.6) and Drosophila melanogaster (29.9) show the lowest betweenness values.

***Clustering coefficient***. The highest clustering coefficient value is observed for Arabidopsis thaliana (0.680) followed by Gallus gallus (0.646) and Saccharomyces cerevisiae (0.607). All the remaining organisms show clustering coefficient < 0.600 suggesting that, in the smaller networks, proteins tend to be directly connected with their neighbors.

***Communities and Modularity***. The modularity and number of communities provide a different description of alteration in a network, since it is an evaluation based on its configuration models. In **Table 2** we report the number of communities and values of modularity calculated with the Louvain algorithm. Mus musculus and Saccharomyces cerevisiae have the same number of communities (6) and the same modularity (0.352). Caenorhabditis elegans has the highest modularity (0.450), together with Macaca fascicularis (0.419) although the number of communities is different (6 and 5 respectively). Gallus gallus provides the smallest number of communities (2), followed by Drosophila melanogaster and Arabidopsis thaliana (3), probably because the sizes of the networks are quite smaller and comparable among them, concerning the other organisms (29, 57 and 62 nodes). A graph representation of main communities for all the native networks is shown in **Figure 3**.

***Correlation Analysis***. Pairwise comparisons for degree, closeness and betweenness are evaluated via Pearson correlation coefficient. This analysis is performed only across these quantities because they are the more common centrality measures. Betweenness centrality increases with the degree ($r = 0.65$; $p = 0.02$) but it shows a strong negative correlation with the closeness ($r = -0.78$; $p = 0.003$). Degree and closeness centrality show a non-significant moderate negative correlation ($r = -0.43$; $p = 0.16$), indicating opposite behavior for these measures.

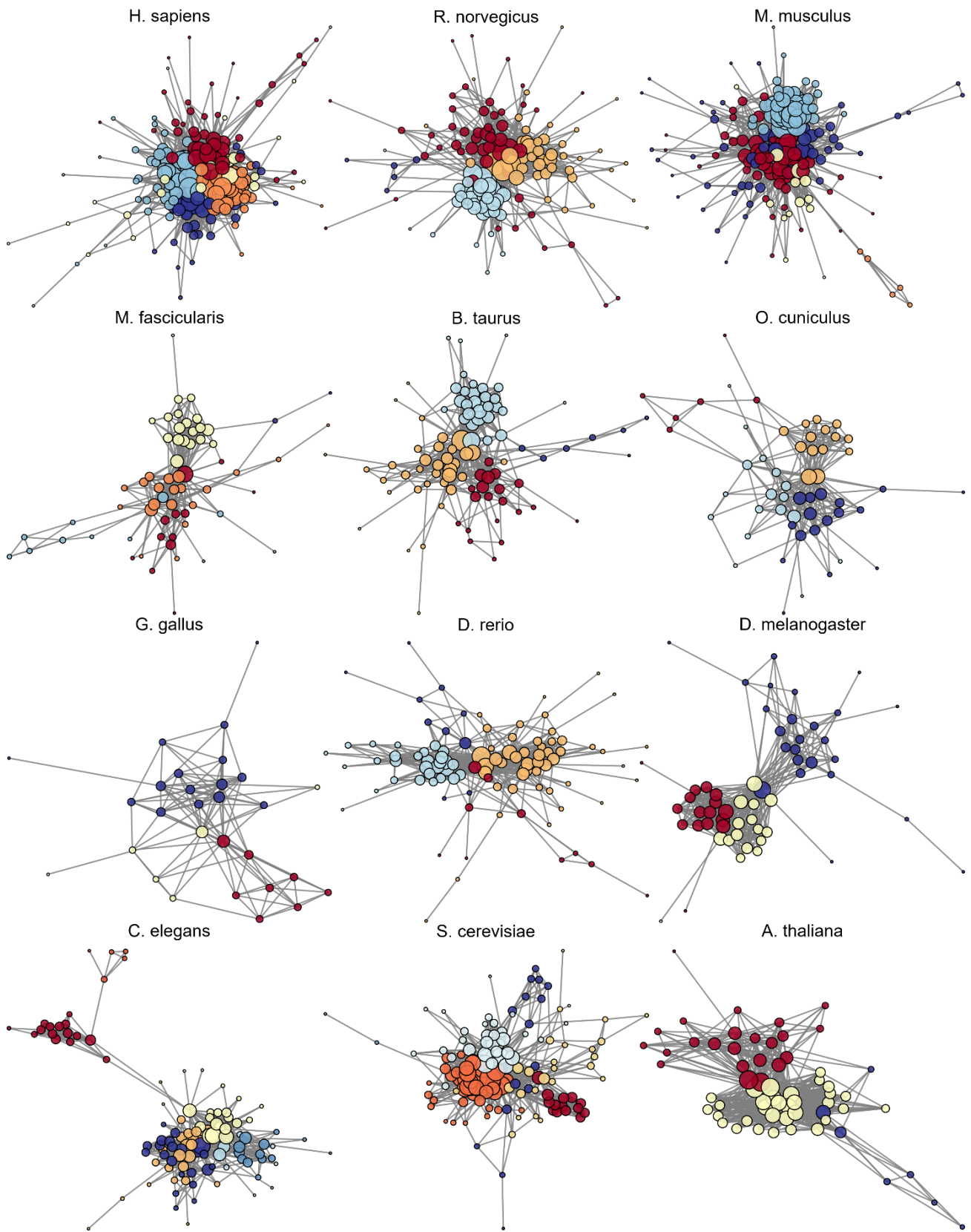

*Figure 2: Graph representation of native UPR models across species. Distinct colors identify communities, and the size of the nodes is related to their degree.*

*Normalized Network Characteristics*

The native UPR models for the different organisms are characterized by different sizes that may affect the standard measures reported in the previous paragraph. Thus, when analyzing normalized metrics (based on the network size), analogies and differences between organisms becomes clearer (**Table 3**).

*Table 3: Normalized values and statistical analyses of common network features for all native models.*

| | Degree | Closeness | | | Betweenness | | | Clust. coeff |
|---|---|---|---|---|---|---|---|---|
| | Norm. value | Norm. value | z-score | p-value | Norm. value | z-score | p-value | z-score |
| Homo sapiens | 0.116 | 0.440 | -3.0 | ≪ 0.05 | 0.063 | 1.9 | < 0.05 | 3.0 |
| Rattus norvegicus | 0.115 | 0.434 | -3.0 | ≪ 0.05 | 0.043 | -1.1 | 0.50 | 3.0 |
| Mus musculus | 0.110 | 0.432 | -3.0 | < 0.05 | 0.042 | -1.1 | 0.50 | 3.0 |
| Macaca fascicularis | 0.141 | 0.424 | -3.0 | ≪ 0.05 | 0.080 | -0.5 | 0.10 | 3.0 |
| Bos taurus | 0.129 | 0.433 | -3.0 | ≪ 0.05 | 0.053 | -0.9 | > 0.10 | 3.0 |
| Oryctologus cuniculis | 0.185 | 0.458 | -2.9 | ≪ 0.05 | 0.012 | 1.1 | > 0.50 | 3.0 |
| Gallus gallus | 0.310 | 0.556 | -2.3 | > 0.10 | 0.130 | -1.1 | 0.50 | 2.9 |
| Danio rerio | 0.113 | 0.401 | -2.9 | ≪ 0.05 | 0.083 | 0.2 | > 0.10 | 3.0 |
| Drosophila melanogaster | 0.246 | 0.501 | -2.0 | < 0.05 | 0.070 | -1.6 | > 0.50 | 3.0 |
| Caenorhabditis elegans | 0.101 | 0.368 | -3.0 | ≪ 0.05 | 0.048 | -2.7 | < 0.05 | 3.0 |
| Saccharomyces cerevisiae | 0.127 | 0.422 | -3.0 | ≪ 0.05 | 0.092 | -1.3 | < 0.05 | 3.0 |
| Arabidopsis thaliana | 0.306 | 0.563 | -3.0 | ≪ 0.05 | 0.135 | 0.7 | > 0.50 | 3.0 |

***Average Degree***. Three mammals (i.e. Homo sapiens, Rattus norvegicus and Mus musculus) share similar values of degree (0.116, 0.115 and 0.110), together with the Danio rerio (0.113). The Caenorhabditis elegans model provides the lowest values of normalized degree (0.101) while Gallus gallus and Arabidopsis thaliana have the highest degree values (0.310 and 0.306). Yet, the Oryctologus cuniculis shows a value of normalized degree (0.185) between the variation range.

***Closeness centrality***. Homo sapiens, Rattus norvegicus, Mus musculus and Bos taurus share also similar values of closeness centrality (0.440, 0.434, 0.432 and 0.433). Caenorhabditis elegans model provides the lowest values of closeness centrality (0.368) while Gallus gallus, Arabidopsis thaliana and Drosophila melanogaster have the highest closeness values (0.556, 0.563 and 0.501).

***Betweenness centrality***. Rattus norvegicus, Mus musculus and Caenorhabditis elegans share comparable values of normalized betweenness (0.043, 0.042 and 0.048). Gallus gallus and Arabidopsis thaliana show the highest betweenness values >0.1 (0.130 and 0.135, respectively).

***Configuration models and z-score distributions***. To assess concrete differences between organisms models, we resort to the use of configuration models. For each real-world UPR model we reconstruct 10 configuration models, to compute the z-scores of closeness centrality, betweenness centrality and clustering coefficient, reported in **Table 3**. Mus musculus and Rattus norvegicus show same values of z-scores for all the quantities, while similarities with Homo sapiens model are observed only for closeness centrality and clustering coefficient. Drosophila melanogaster and Saccharomyces cerevisiae have comparable z-score for closeness centrality while Caenorhabditis elegans and Saccharomyces cerevisiae show similar z-score for betweenness centrality. Overall, z-scores of closeness and clustering coefficient are included in small variation ranges ([-3.0, -2.3] and [2.9, 3.0] respectively). Moreover, z-score for betweenness centrality shows significant differences among organisms (from -2.7 for Caenorhabditis elegans to 1.9 for Homo sapiens), suggesting that betweenness centrality provides a potential useful tool for identifying similarity or differences between organisms regarding this specific mechanism.

***Configuration models and statistical analysis***. In **Table 3**, we also report results from nonparametric signed rank test for closeness and betweenness centralities. Regarding closeness centrality, we obtain that the native and the configuration models are significantly different for all species, except for Gallus gallus ($p > 0.10$). In fact, all the p-values for closeness are smaller (or much less) than 0.05, so the test rejects the null hypothesis of zero median at the 5% significance level. On the other hand, for the betweenness centrality distribution, the test provides a significant difference only for half of the sample (Homo sapiens, Caenorhabditis elegans, and Saccharomyces cerevisiae), but distributions cannot be distinguished in the rest of the organisms. Overall, the configuration models alter the network features regarding closeness and only partially regarding the betweenness centrality.

***Highest metrics nodes***. To more precisely relate the evaluated network metrics with the biological content, we also analyze the role of protein within the UPR pathways. Few specific genes or their homologues appear in all the sets across organisms, associated with high values of degree, closeness and betweenness centralities. In **Table 4** we report the nodes with the highest values of the three principal metrics.

A percentage varying between the 5% and 30% of the total number of proteins in each native network is represented by heat shock cognate proteins (HSC), that are members of the heat shock protein family (HSP), one of the most ubiquitous and conserved protein families across organisms [79–82]. They are fundamental in the correct functioning of cells, maintaining the cellular proteostasis and protecting cells from induced stresses [83,84]. Their genes are associated to the highest values of the main three network features, across all the twelve species. They are found as relevant nodes and hubs also in various diseases, like cancers and strokes [85–88]. HSP/HSC proteins for each organism belong to the same community, the one associated with the largest node size (**Figure 2**). Other proteins related to the highest values of the three metrics for the various networks are: i) X-box binding protein 1 (XBP1), which is an important initiator and modulator factor of ER stress response [89,90], ii) ATF6 and iii) ERN1, both of which are ER membrane receptors (together with EIF2AK3) [9,91–93], in charge of initiating and regulating the stress response after the activation promoted by HSPA5/BiP [94–96]. Moreover, focusing on the proteins that can be associated to the main subnetwork of UPR [97] (**Table 4**), another main result is that the most relevant pathway in the UPR mechanism for Saccharomyces cerevisiae and Arabidopsis thaliana is related to the IRE1/ERN1 signaling cascade [98–103], also notable from the fact that we cannot find homologues for the other two ER stress sensors.

*Table 4: Nodes sets with highest values of network metrics, for native models. Proteins' homologues are found in the Homologous Gene Database (HGD, https://ngdc.cncb.ac.cn/hgd) [104] and UniProt database [105–107]. Not all proteins have a corresponding homologous in our reference minimal network model [97].*

| | Degree | Closeness | Betweenness | Homologous of UPR proteins from Ref. [97] |
|---|---|---|---|---|
| Homo sapiens | HSPA5, HSPA4, HSP90AB1, HSP90AA1, HSPA8 | HSPA5, HSPA4, HSP90AB1, HSP90AA1, HSPA8 | HSPA8, HSP90B1, XBP1, HSP90AA1, HSP90AB1, ATF6 | ATF4, ATF6, DDIT3, EIF2AK3, EIF2S1, ERN1, HSPA5, NFE2L2, PPP1R15A, XBP1 |
| Rattus norvegicus | HSPA5, HSP90AB1, HSP90AA1, HSPA1B, XBP1 | HSPA5, HSP90B1, XBP1, HSPA1B, HSP90AA1, ATF6, DNAJC3 | HSP90B1, XBP1, ATF6, ERN1, HSP90AA1, HSPA1B | Atf4, Atf6b, Ddit3, Eif2ak3, Ern1, Hspa5, Nfe2l2, Ppp1r15a, Xbp1 |
| Mus musculus | HSP90B1, HSPA1B, HSP90AB1, HSP90AA1, XBP1 | HSPA5, HSP90B1, HSPA1B, XBP1, HSP90AB1, HSP90AA1, ATF6 | HSPA1B, CCND1, HSP90B1, ERN1, HSP90AB1 | Atf4, Atf6, Ddit3, Eif2ak3, Ern1, Hspa5, Nfe2l2, Ppp1r15a, Xbp1 |
| Macaca fascicularis | HSPA5, HSP90B1, XBP1, DNAJC3, ATF6, HSPA1L | HSPA5, XBP1, HSP90B1, DNAJC3, ATF6, CALR, ATF6B | HSPA5, XBP1, CALR, HSP90B1, DNJC3, ATF6 | ATF4, ATF6, DDIT3, EIF2AK3, EIF2S1, ERN1, HSPA5, NFE2L2, PPP1R15A, XBP1 |
| Bos taurus | HSPA5, HSP90B1, DNAJC3, ATF6, XBP1, HSP90AA1, HSPA9 | HSPA5, HSP90B1, DNAJC3, ATF6, XBP1, CALR, DERL1 | HSPA5, DNAJC3, HSP90B1, ATF6, DERL1, TMED2 | ATF4, ATF6, DDIT3, EIF2AK3, EIF2S1, ERN1, HSPA5, NFE2L2, PPP1R15A, XBP1 |
| Oryctolagus cuniculus | HSPA5, HSP90B1, XBP1, DNAJC3, ATF6, CALR | HSPA5, HSP90B1, XBP1, DNAJC3, ATF6, EIF2S1, CALR | HSPA5, HSP90B1, XBP1, RNASEL, ATF6, ERN1 | ATF4, ATF6, DDIT3, EIF2AK3, EIF2S1, ERN1, HSPA5, NFE2L2, PPP1R15A, XBP1 |
| Gallus gallus | HSPA5, HSP90B1, HSP90AB1, HSP90AA1, DNAJB1 | HSPA5, HSP90(B1, AB1, AA1), DNAJ(B1, C3, A2) | HSPA5, HSP90B1, DNAJB1, HSP90AB1, CRYAB, HSP90AA1, DNAJC3 | ATF4, ATF6, DDIT3, EIF2AK3, EIF2S1, ERN1, HSPA5, NFE2L2, PPP1R15A, XBP1 |
| Danio rerio | hspa5, hsp90b1, hspa9, sec63, hsp90aa1.1, xbp1, hsp90ab1, hsp90aa1.2, dnajc3a | hspa5, hsp90b1, sec63, hspa9, dnajc3a, canx, xbp1, hsp90aa1.1, edem1, dnajc10 | hspa5, hsp90b1, sec63, txnl1, xbp1, eprs1, edem1, hspa9 | atf4, atf6, ddit3, eif2ak3, eif2s1b, ern1, hspa5, nfe2l2a, ppp1r15a, xbp1 |
| Drosophila melanogaster | HSC70-3, GP93, HSP83, HSP70AB, HSP70BB, HSC70-5, DROJ2 | HSC70-3, GP93, HSP83, DROJ2, CG2918, HSC70-5 | HSC70-3, XBP1, GP93, HSP83 | crc, Atf6, -, EIF2AK3, Ire1, Hsc70-3, cnc, PPP1R15, Xbp1 |
| Caenorhabditis elegans | HSP-70, HSP-90, HSP-4, HSP-3, HSP-6, ENPL-1 | HSP-90, HSP-70, HSP-4, HSP-3, HSP-6, ENPL-1, XBP-1 | HSP-90, HSP-70, PQN-91, DAF-16, HSP-4, HSP-4 ENPL-1, HSP-3, HSP-6 | Atf-4, atf-6, -, pek-1, ire-1, hsp-4, sknr-1, -, xbp-1 |
| Saccharomyces cerevisiae | HSC82, HSP82, KAR2, LHS1, SSA2, SSA1 | KAR2, HSC82, HSP82, SSA1, SSB1, SSA2, HSP104, SSB2 | ATG8, HSC82, SSA1, SSB1, SIL1, IRE1, HAC1, CDC48, SSB2 | MET4/MET28, -, -, ISR1, IRE1, KAR2, - , -, Xbp1 |
| Arabidopsis thaliana | BIP(3, 2), HSP70-4, HSP90-(1-4, 7) | BIP(3, 2), HSP70-4, HSP90-(1-4, 7) | BIP(3, 2), HSP70-4, HSP90-7, CPN60, CCT6A, BZIP60 | -, -, -, -, -, IRE1(A, B), BIP(2, 3), -, -, BZIP(18, 34, 60, 61) |

***Multi-comparison test.*** In **Figure 3** we report the correlation matrices related to multiple comparison test applied to normalized metrics distributions (degree, closeness centrality and betweenness centrality) cross-species. Representation of metrics distributions are shown in **Figure S5** of Supplementary material. We observe that degree and closeness centrality provide similar results. On the contrary, correlation matrix of betweenness centrality shows that most of organisms have very similar normalized distributions (most of matrix elements are closer to 1), so this metrics does not discriminate among all organisms.

We can conclude that statistical and correlation analyses show that closeness centrality, together with the degree, results in a better network feature to discriminate among the same organism and across different organisms.

***Topological analysis***. A different type of information comes from the topological analysis of adjacency matrices, evaluated with the Generalized Hamming Distance (GHD, **Table 5**). It provides a degree of difference between two NxN matrices, by comparing paired matrix elements [49]. In **Table 5** we report the average value of GHD calculated for the native network with respect to each of the 10 associated configuration models. The most different models are provided by Arabidopsis thaliana (0.2559), Gallus gallus (0.2414) and Drosophila melanogaster (0.2143); this can be rationalized because the size of the network is small compared to the other organisms, and we have poor biological information about the nodes, so the null models generate quite different related networks. The most similar matrices are obtained for Mus musculus, and it provides a similar result to Rattus norvegicus (0.1359 and 0.1376). Other comparable GHD values are for Homo sapiens, Saccharomyces cerevisiae and Danio rerio (0.1439, 0.1457 and 0.1422).

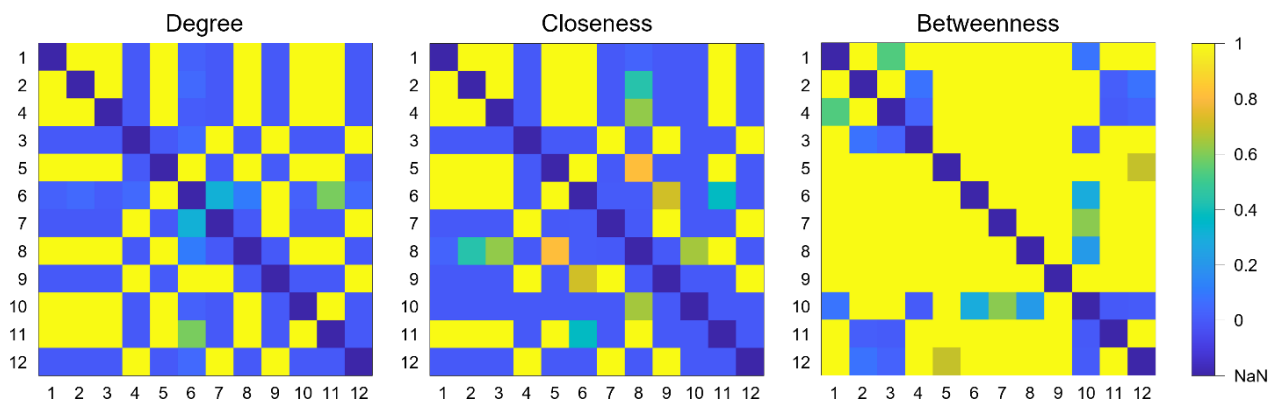

Figure 3: Correlation matrices (p-value) of normalized metrics distributions. Matrix indices represent the organisms as listed in the text. Color scale is the same for all the matrices (color bar on the right). NaN elements identify the diagonal of the matrices (same organism).

Table 5: Average Generalized Hamming Distance between native and configuration models.

| Homo sapiens | Rattus norvegicus | Mus musculus | Macaca fascicularis | Bos taurus | Oryctologus cuniculis | Gallus gallus | Danio rerio | Drosophila melanogaster | Caenorhabditis elegans | Saccharomyces cerevisiae | Arabidopsis thaliana |
|---|---|---|---|---|---|---|---|---|---|---|---|
| 0.1439 | 0.1376 | 0.1359 | 0.1617 | 0.1519 | 0.1873 | 0.2414 | 0.1422 | 0.2143 | 0.1397 | 0.1457 | 0.2559 |

We can conclude, from statistical and correlation analyses, that closeness centrality, together with the degree, are convenient network features, able to discriminate across different organisms, and to identify native or null networks within the same organism.

Summarizing the results for all the above quantities, the mammals (Homo sapiens, Rattus norvegicus, Mus musculus and Bos taurus) and their biological models (Danio rerio, Caenorhabditis elegans and Saccharomyces cerevisiae) share similar values, in particular of closeness, and have similar connections, as shown by the correlation matrices and the GHD values.

Overall, the description of UPR networks via standard network quantities and GHD allows to identify phylogenetic similarities, and to characterize the networks of mammals with respect to other phyla.

**b. Network robustness**

The networks robustness (**Figure 4**) has been tested both with random- and metrics-based target attacks to the native networks. In **Table 6** we report normalized values of the average path length and efficiency for all investigated network models. Average shortest path length is observed for two of the smallest network models (1.840 and 1.860 for chicken and plant, respectively), with Caenorhabditis elegans providing the highest value (2.832). As expected, the opposite situation is reflected in the normalized global efficiency of the networks, since higher values are related to more robust networks. Comparable values of robustness quantifiers are obtained for the five mammals (~2.3-2.4 and ~0.48 for human, murine, monkey and the bull), together with the yeast. Another comparable couple is provided by the rabbit and the fruit fly (~2.1-2.2 and ~0.5).

Table 6: Normalized values of average path length and efficiency of all native models.

| | Average path length | Efficiency |
|---|---|---|
| Homo sapiens | 2.354 | 0.488 |
| Rattus norvegicus | 2.382 | 0.483 |
| Mus musculus | 2.393 | 0.479 |
| Macaca fascicularis | 2.453 | 0.486 |
| Bos taurus | 2.396 | 0.487 |
| Oryctologus cuniculis | 2.277 | 0.523 |
| Gallus gallus | 1.860 | 0.629 |
| Danio rerio | 2.601 | 0.456 |
| Drosophila melanogaster | 2.068 | 0.571 |
| Caenorhabditis elegans | 2.832 | 0.429 |
| Saccharomyces cerevisiae | 2.450 | 0.478 |
| Arabidopsis thaliana | 1.840 | 0.633 |

The evolution of the largest connected component (LCC) in each network is evaluated by removing at each step one node from the network based on ascending index (random [108], degree [109] and centralities [110] attacks, i.e. the least "important" nodes are removed first. The LCC identifies a connected component of a given graph that contains a significant fraction of the entire graph's vertices. The removal based on random node choice provides a linear trend for all organisms (**Figure 4**). An attack strategy relying on random removal of nodes requires the removal of many nodes for significantly decreasing the potency of the attack, so targeted attacks result more efficient in degrading the network [111,112]. In most cases, the betweenness-based attack strongly affects the behavior of the network, except for Caenorhabditis elegans, for which the opposite is observed. In general, degree-based and closeness-based attacks show the same behavior of LCC degradation, except for Macaca fascicularis, Bos taurus and Arabidopsis thaliana which provide similar trends for the two centralities attacks. Gallus gallus provides a situation in which all the three metrics-based attacks have comparable effects on the robustness of the network model.

Overall, degree and closeness are quite correlated features in this type of analysis. This finding is also supported by sets of nodes with highest values of metrics (see **Table 4**), where it is shown that sets for degree and closeness are more similar compared to sets obtained for betweenness centrality, as confirmed also by the computed correlation coefficients between network metrics (see above). Interestingly, Macaca fascicularis, Bos taurus, Danio rerio and Arabidopsis thaliana show a degree-based degradation in the robustness quite similar to the random removal, so much slower compared to the two centrality-based attacks.

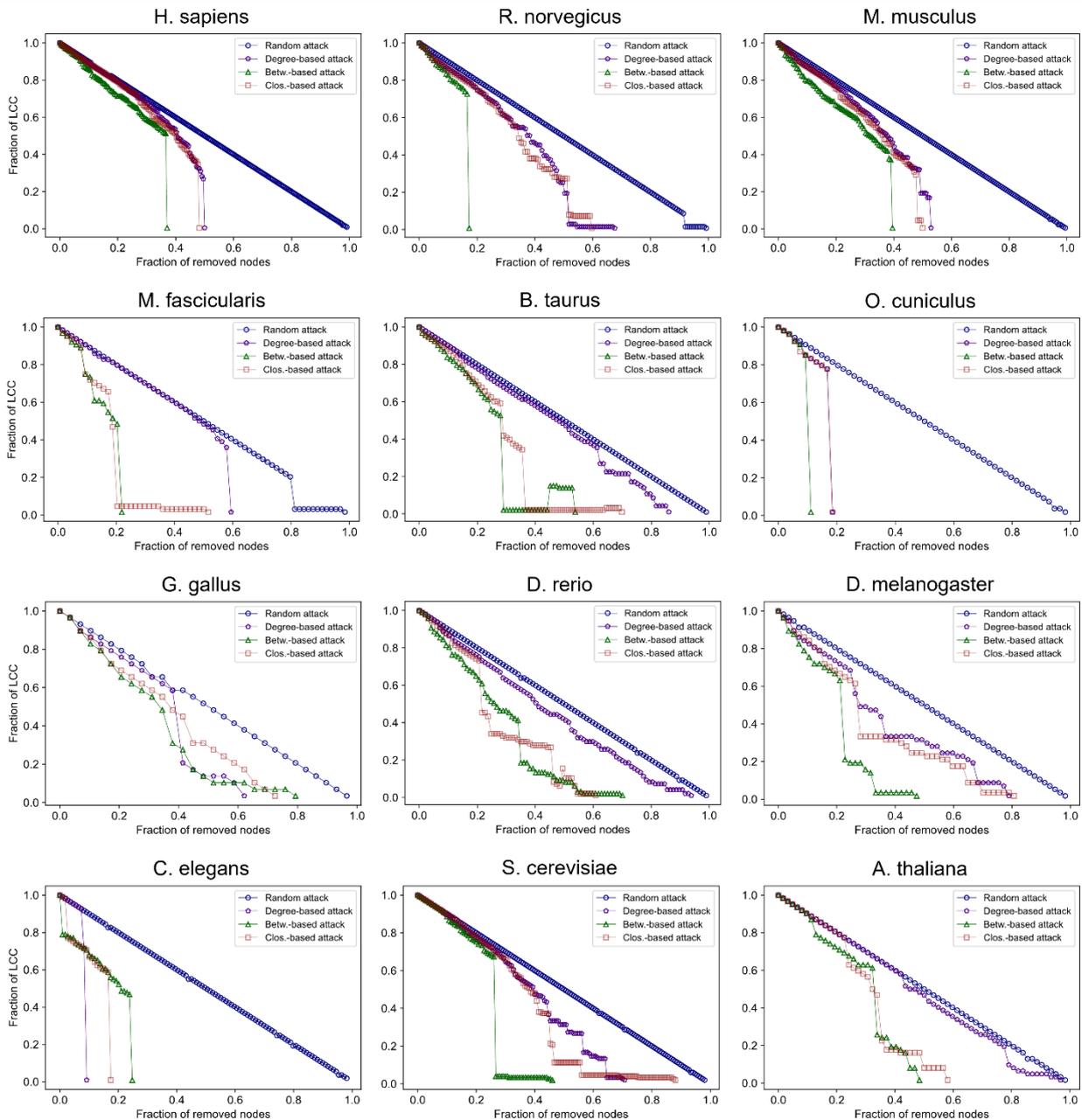

*Figure 4: Network robustness evaluated using random attack and various target attacks on native models.*

By removing nodes with increasing value of metrics, we expect a slower degradation of the networks integrity (notable especially for the human, murine, bull, zebrafish, yeast and plant models); the results could be explained considering a larger number of "external" and less important nodes within the networks (very low values of network metrics). Attacking the networks based on decreasing values of metrics provides a random-like trend in network robustness.

The sudden jumps in the robustness can be associated not only to a precise role of a specific protein removal, because their removal at the beginning of the process could not produce the same results, like in a purely star network. Instead, it can also be the result of a cumulative effect, because the overall removal of all the previous proteins builds a star-like network, that collapses upon removal of a specific protein.

However, going through the principal functions and pathways that are involved in the UPR and arising from the network robustness, across all, or most of, the organisms, we highlight that those proteins with the role of chaperones, i.e. initiating the signaling pathways, are present in all considered organisms (apart Caenorhabditis elegans) as the ones that induce a sudden collapse of metrics robustness. Also, proteins involved in pro-apoptosis and ERAD mechanisms emerge from the robustness analysis as present in most organisms. Other mechanisms, with corresponding proteins as obtained from robustness stress analysis, are listed in **Table 7**.

*Table 7: UPR key mechanisms, pathways and functions sustaining the network robustness.*

| | Chaperone | TM receptor and transfer | Kinase | Regulator and messenger | Pro-apoptotic and ERAD | Pro-survival | Initiation and transcription factor | Protein folding | Nonspecific in UPR |
|---|---|---|---|---|---|---|---|---|---|
| Homo sapiens | HSPD1 | – | – | – | UBQLN1, TMBIM6 | – | – | – | – |
| Rattus norvegicus | Hspb2 | – | – | – | Tomm20 | – | – | Erp44 | – |
| Mus musculus | Hspb6, Dnajb12 | – | – | – | Vapb, Os9 | – | Creb3, Creb3l1 | – | Cops5 |
| Macaca fascicularis | CRYAA | – | – | – | BAG6, HERPUD2 | – | CREBRF | – | – |
| Bos taurus | HSPB1, DNAJA2, DNAJC10 | TMTC4 | ERN2 | – | ERN2, DNAJC10, HERPUD2 | – | – | – | – |
| Oryctologus cuniculis | HSPA5 | – | – | – | RNASEL | – | – | – | – |
| Gallus gallus | DNAJA4, DNAJB1, DNAJA3 | ERN1 | – | – | – | – | – | – | – |
| Danio rerio | hsp90aa1.2, dnajc10 | PEK | – | – | dnajc10, heprud2 | ddrkg1 | – | – | – |
| Drosophila melanogaster | CG4461, CG5504, Hsp67Ba | – | – | – | Edem2, Der-1 | Ficd, wfs1 | – | – | – |
| Caenorhabditis elegans | – | – | – | – | – | – | pqn-2, nhr-80, pqn-79, pqn-90 | – | set-6 |
| Saccharomyces cerevisiae | APJ1, GIM3 | – | – | – | – | – | BTT1 | – | – |
| Arabidopsis thaliana | PFD6, DJA5, HSP70-6, HSP90-3, HSP70-4 | F26K24.16 | – | F23H11.4 | – | – | BZIP60 | – | – |

### c. Structural controllability

The structural controllability allows to determine the number and identity of the minimum driver nodes (**Table 8**). For all organisms, the study of the native model may produce different number of driver nodes sets, and of driver nodes.

For Homo sapiens, Gallus gallus, and Drosophila melanogaster the algorithm finds 1 single set of driver nodes (namely, of RPA2P, HSP70AB/I(2)EFL, and STT3B).

For all other organisms, the algorithm finds different possible driver nodes sets, moreover, for the same organisms, few nodes occur in all the possible allowed driver nodes sets. Recurrent proteins in the driver nodes sets are: UFL1 for Rattus norvegicus, Ficd for Mus musculus, CRYAB/NCK1/VAPB for Macaca fascicularis, abu-12 for Caenorhabditis elegans, and NAC062/PFD1 for Arabidopsis thaliana. The algorithm cannot find driver nodes sets for Bos taurus, Oryctologus cuniculis and Danio rerio. This is not surprising, because it is possible that the description of these networks lacks connections necessary to define the scheme of structural controllability.

***Gene analysis of driver nodes.*** We provide some details on genes identified as driver nodes, to identify general trends and behaviors in the different networks.

| | DNs sets | # DNs sets |
|---|---|---|
| Homo sapiens | 1 (RPAP2) | 1 |
| Rattus norvegicus | 2 (UFL1, CREB3L4 or CREB3L2) | 2 |
| Mus musculus | 4 | 8 |
| Macaca fascicularis | 5 | 4 |
| Bos taurus | – | – |
| Oryctologus cuniculis | – | – |
| Gallus gallus | 1 (STT3B) | 1 |
| Danio rerio | – | – |
| Drosophila melanogaster | 2 (HSP70AB, I(2)EFL) | 1 |
| Caenorhabditis elegans | 3 | 3 |
| Saccharomyces cerevisiae | 4 | 16 |
| Arabidopsis thaliana | 9 | 8 |

The RPAP2 gene, associated with transcription and RNA processing, connects to RUBVL2 and has minimal network metrics [113]. In metrics terms, this node is characterized by 1 degree (connected to RUBVL2), $1.48 \cdot 10^{-3}$ closeness and 0 betweenness (we can define it an "external node").

In Rattus norvegicus, the two alternate driver nodes (CREB3L family of transcription factors), in combination with UFL1, are homologues and they possess distinctive transcriptional activities through their binding to cAMP response elements [114].
The FIC domain protein adenylyltransferase (FICD) is an enzyme belonging to the Fic (filamentation induced by cAMP) domain family. FICD is associated with various cellular pathways, particularly the ATF6 and EIF2AK3 branches of the UPR pathway, which regulate ER homeostasis. In humans, FICD is typically present at very low basal levels in most cell types, and its expression is tightly regulated [115,116].

In mammals, STT3B facilitates the degradation of misfolded proteins via the ERAD pathway [117]. The HSP70AB gene in fruit flies, part of the Hsp70 family, stabilizes non-native protein conformations [118–122]. The I(2)EFLS gene in Drosophila [123] induces phosphorylation of eIF2α, playing a role in aging [124–127]. The Abu-12 gene in C. elegans encodes UPR proteins that function alongside the canonical UPR pathway in xbp-1 mutants under ER stress [128–130].

In addition, we also analyze the altered behavior of the native models by excluding from the networks the DNs obtained in the first round of analysis. The algorithm is not able to find other combinations of driver nodes if the original identified nodes are excluded from the networks.

Control theory applied to UPR networks identifies driver nodes related to common molecular pathways. In rat, two driver nodes are transcription factors; in fruit fly, two are chaperones. The gene Ficd is present as jump node in Drosophila melanogaster robustness and as driver node in mouse and pertains to two of the main pathways of signaling. Macaca fascicularis provides the chaperone CRYAB as recurrent driver node, together with VAPB that also results as jump node for Mus musculus, and NCK1 that is an ERN1 regulator within its signaling pathway. PFD1 protein of Arabidopsis thaliana is another molecular chaperone, like PFD6 which arises from the network robustness.
Overall, it is not possible to identify one single mechanism or protein that is always present across all or most organisms, and this could be due to missing information and/or intrinsic network differences, but some genes and/or mechanisms are recurrent as in the robustness analysis.

## Discussion

Unfolded Protein Response stands as one of the highly conserved fundamental biological mechanisms in organisms [15,16,131,132], occurring primarily in the endoplasmic reticulum (ER). Its main function is to restore cell homeostasis after a pathological accumulation of non-properly folded proteins [6,7,10,11,133,134]. Any inefficiency in the adaptive response to ER stress can lead to the accumulation of unfolded or misfolded proteins at different levels. These proteins tend to aggregate, posing a threat to cellular and tissue integrity and serving as a primary driver for the onset of amyloidosis and neurodegenerative diseases [25–27,135].

A key signaling pathway governing the UPR, originally discovered in Saccharomyces cerevisiae during the 1970s [52,136–138], is characterized by a single transmembrane protein, ERN1, responsible for detection of ER stress provoked by over-accumulation of unfolded/misfolded proteins [98–100]. The major ER chaperone BiP triggers the dimerization of ERN1, which leads to its subsequent autophosphorylation and the activation of its signaling cascade [9]. This pathway reinforces the ER function, and it is conserved across eukaryotes [139]. Indeed, the basic features of the UPR mechanism result to be highly preserved throughout metazoans;

most species have homologues of the three main stress sensors, ERN1, ATF6 and EIF2AK3. Signaling pathways of the stress sensors cooperate to restore and/or bolster ER function, primarily through upregulation of many components of the protein folding machinery (as the action of XBP1 on regulation of BiP chaperone [89,90] within the ERN1 pathway) and the quality control machinery within the ER. Additionally, these signaling pathways work to limit ER stress by dampening the attenuation of translation and potentially engaging the regulated IRE1-dependent decay (known as RIDD) [140–142].

A network-based description of cell mechanisms, using protein-protein interactions (PPI) networks, offers a valuable tool for comprehending the behavior of complex systems. Some points must be considered before discussing the analysis results. PPI network models are built utilizing biological data sourced from publicly available PPI databases. These databases collect various types of data, from thousands of experimental works. Despite their current large size, the databases are not exhaustive, because only part of the molecular pathways have been completely understood and characterized. The absence or bias in information stored in interaction databases must be considered when creating and analyzing biological networks, and PPI databases must contain sufficient information for a specific pathway to yield qualitative accurate results from network analysis techniques.

With these premises, we apply a classical network metrics investigation to assess the degree of similarity and differences existing among UPR models from twelve different organisms, considering the existing phylogenetic pathway in all the analyzed models. As model organisms, we choose for our investigation Homo sapiens, Rattus norvegicus and Mus musculus due to the quite complete genomic accordance with human [143–148], and other model organisms, as Macaca fascicularis [58], Bos taurus [57,59], Oryctologus cuniculis [60,61], Gallus gallus [62,63], Danio rerio [65,66], Drosophila melanogaster [149,150], Caenorhabditis elegans [151], Saccharomyces cerevisiae [152], and Arabidopsis thaliana [76,77,103].

Comparisons made for common metrics provide that network measures are very sensitive to the network features (**Table 1**, and **Figures S2–S4** from Supplementary); however, from a biological point of view, the network analysis shows that there are strong similarities among the model organisms (**Table 3**), in particular among mammals and their model organisms. Moreover, there are highly conserved genes and pathways across species (**Table 4**). Chaperones belonging to the heat shock protein (HSP) family are among the most central proteins in terms of network metrics, as the densest connected, and they also play a fundamental role in the correct functioning of cells [79–83]. This result can be interpreted as a support to significance of these proteins, as it implies a correlation between the structure and the biological properties of protein networks. In summary, network theory and statistical mechanics confirm that it is possible to identify similarities among organisms phylogenetically related via their PPI networks. In addition, we find that there are some network metrics that better discriminate among organisms, in the sense that are useful for identifying similarities and differences, as closeness centrality and node degree.

Additionally, in this study we employ a network scientific methodology to investigate the levels of tolerance of multiple systems when subjected to external perturbations. We achieve this by adapting a measure of network robustness [153–156], to characterize the potential resilience of several PPI networks. The levels of robustness of the twelve models are evaluated by removing nodes, and consequently altering the network integrity, adopting different network-based metrics target strategies. Obviously, in these cases, network robustness is influenced by network features and degree of accuracy of biological information accessible online, but similarities in resilience behavior arise for organisms that are phylogenetically closer, for example Homo sapiens and murine species, despite the difference between network features. This result can be supported by the closeness between these organisms in terms of phylogeny, since 99% of the genome is conserved between human and murine species [143–148]. Similar behaviors can be also found based on network-size – i.e., bovine, plant and machaca, or zebrafish and fruit fly. A detailed analysis of genes that are related to jumps in the network robustness highlights that, despite different proteins are involved in different organisms, there is a recurrence of pathways across most species. In particular, the sensing role of the chaperones is fundamental, together with the apoptosis and endoplasmic-reticulum-associated protein degradation (ERAD) functions.

Lastly, the minimum drive nodes methodology, based on the structural controllability analysis [41,42] and employing Kalman's rank condition [157], yields significant biological insights into the proteins involved in specific mechanisms. Through the application of this theory to network models, it becomes feasible to pinpoint the key nodes that exert control over the entire network [41,42,158]. We apply structural controllability to all organism models investigated in this work. Driver nodes sets are related to the topological structure of the network; as for the other network methods applied, results are sensitive to the biological information available online (also resulted in the high variability of the possible DNs sets obtained – several sets for the same organism in **Table 8**). Attributing a biological significance to the set of driver nodes is not straightforward. When analyzing networks with a high number of nodes and edges, there is a possibility of losing biological information in the models due to the consequent increase of missing information stored in the databases. However, some specific mechanisms and genes, already identified via robustness and network statistics, arise also in the list of driver nodes.

This study represents a comprehensive and innovative analysis of the biological behavior and characteristics of a fundamental cellular control mechanism, the Unfolded Protein Response (UPR). By modeling the UPR as a protein-protein interaction (PPI) network, it uses both standard and advanced techniques to extract meaningful information from the network. The study establishes a direct correlation between specific network features and biological components. It finds that the three classes of methods employed – standard network metrics and topology, network robustness, and network control theory – offer complementary and non-conflicting characterizations of the systems studied.

The first class of methods offers a comprehensive portrayal of the PPI networks and clearly discerns phylogenetic similarities, distinguishing mammals and their animal models (Danio rerio, Caenorhabditis elegans, and Saccharomyces cerevisiae) from organisms of other phyla. Additionally, it identifies key genes (network nodes) that significantly influence the overall UPR mechanism. Network robustness utilizes network metrics to assess the network's ability to withstand node removal, identifying pivotal proteins and sub-pathways, and simulating disease onset. This robustness analysis also highlights phylogenetic similarities among organisms.

Among the various UPR sub-mechanisms, chaperone sensing, apoptosis, and ERAD are identified as the most relevant. The control theory, though less distinctly, also pinpoints proteins with central roles, particularly chaperones, transcription factors, and ERAD proteins. Using network models for molecular and cellular pathways is a powerful yet underutilized approach. This study's combination of diverse network descriptors and methods provides profound insights into complex mechanisms and emphasizes the need to revise and enrich PPI databases to create increasingly accurate biological models.

## Materials and Methods

In this work, we present a study of the unfolded protein response mechanism among twelve organisms.

Below we report general information of the network models investigated in this study, and the description of the performed analyses. We perform calculation of i) network descriptors: degree, betweenness and closeness centralities, ii) modularity and communities, iii) topological distance, iv) network robustness, and v) structural controllability. We perform comparisons among native models, with the help of the configuration models (degree-based reconstruction). Native models are built using Python v. 3.11 programming language and NetworkX Python library [159], based on information on paired connection between couples of nodes, and configuration models are built using MATLAB v. R2023a programming language. Modularity and communities, and robustness analyses are performed using NetworkX library, while network descriptors analysis and structural controllability [160,161] have been implemented in MATLAB.

### UPR network models

Here we present the UPR network models proposed in this work. Investigated organisms are: Homo sapiens, Rattus norvegicus, Mus musculus, Macaca fascicularis, Bos taurus, Oryctolagus cuniculis, Gallus gallus, Danio rerio, Drosophila melanogaster, Caenorhabditis elegans, Saccharomyces cerevisiae, and Arabidopsis thaliana. We combine information stored in UniProt database [105–107,162] and String protein-protein interaction database [163–166] to identify proteins involved in the mechanism for all investigated organisms and build undirected network models. We impose a minimum required interaction score of medium confidence, and we consider experimental evidence, curated databases, text-mining and co-expression associations to construct the set of connections among nodes. In this framework, original models, also referred as native, are built directly from biological information obtained by the consulted databases, while connections in configuration models are semi-randomly built based on total degree of each node in the original models. Matrix representation of native UPR models is shown in **Figure S1** of Supplementary material.

### Analysis of network models using network metrics

We first characterize the models using the usual metrics of network descriptors, (i) to highlight differences and similarities between models related to different organisms and (ii) to evaluate the influence of constructing a network with random connections starting from the degree of nodes of original models.

For each model, we calculate (i) the total degree of the nodes, (ii) the betweenness and closeness centralities, which are measures of how often each graph node appears on the shortest path between two nodes in the graph, (iii) local clustering coefficient, and (iv) the modularity and number of communities, as measures of the structure of networks to evaluate the strength of division into different modules, calculated using the Louvain community detection Algorithm. As definition, networks with high modularity have dense connections within communities.

In addition, we also include normalized values of degree ($\mathbf{norm(deg)} = \mathbf{2M/N}$, where 2M is the total degree of a node in a N nodes network) [167], betweenness ($\mathbf{norm(bet)} = (\mathbf{bet} - \mathbf{min(bet)})/(\mathbf{max(bet)} - \mathbf{min(bet)})$) and closeness centrality ($\mathbf{norm(clos)} = (\mathbf{N - 1}) \cdot \mathbf{clos}$) [168] (Table 3). All measures are normalized by removing the dependence from the network dimensionality (normalized distributions shown in Figures S2-S4 of Supplementary material). We compute z-scores of averaged normalized metrics distributions, defined as $x - \mu)/\sigma$, where $x$ is the value of the variable, $\mu$ and $\sigma$ are the mean value and standard deviation of the population, respectively.

### Analysis of topological differences between original and configuration models

As well explained in our previous work [97], we use the algorithm relies on the Generalized Hamming Distance (GHD) [169], which can be used for assigning a "weight" to the topological difference between networks, and evaluating its statistical significance, based on comparison between matrix elements. We apply this theory to evaluate the degree of difference between original models (built from the databases) and the configuration models (re-created from original models).

If we consider two distinct networks, labeled X and Y, with the same number of nodes (N), we can calculate the distance dGHD between the two networks as follows:

$$\mathbf{dGHD(X, Y)} = \frac{1}{N(N-1)} \sum_{i \neq j} (\mathbf{x'}_{ij} - \mathbf{y'}_{ij}) \qquad (1)$$

where $\mathbf{x'}_{ij}$ and $\mathbf{y'}_{ij}$ are mean centered edge-weights, and depend on the topology of the network, providing a measure of connectivity between every pair of i[th] and j[th] nodes in X and Y, respectively.

### Statistical analysis of network models

The graph analysis, together with the network robustness, is performed using Python 3.11 [170]. The statistical analysis, the implementation of structural controllability and the GHD algorithm are performed using MATLAB R2023a [171]. We apply non-parametric tests because our variables are not normally distributed. For paired comparisons between the centralities distributions of native and null models considering one specific organism we use the Wilcoxon signed-rank test, and for non-paired comparisons across different organisms we use a multi comparison test computing values with the Bonferroni method [172–178], on the results of a one-way Kruskal Wallis analysis of variance (shown in Figure S4 of Supplementary material). If p < 0.05, the results are regarded statistically significant.

### Network robustness exploitation

Considering a network X, composed of N nodes set denoted as $\mathbf{V} = \{\mathbf{v_1}, \mathbf{v_2}, ..., \mathbf{v_N}\}$, interconnected by M links represented by $\mathbf{E} = \{(\mathbf{v_i}, \mathbf{v_j}): \mathbf{v_i}, \mathbf{v_j} \in \mathbf{V}\}$, the robustness $\mathbf{R}$ of the X network is defined by the ratio [179,180]

$$\mathbf{R} = \frac{1}{N} \sum_{i=1}^{N} \mathbf{G_i} \qquad (2)$$

where $\mathbf{G_i} = \mathbf{n_i}/\mathbf{N}$ is the size of the large connected component after the removal of i[th] node. The normalization factor $\mathbf{N^{-1}}$ is useful for the comparison of networks of different sizes. To numerically quantify the robustness of a network, some metrics can be used. The average path length $l$ can provide a quantification of network robustness, since large values of $l$ mean that nodes are farther apart from each other, and removal of a node can significantly increase the average paths between many other nodes, decreasing the robustness of the network [181]:

$$l = \frac{1}{N \cdot (N-1)} \sum_{i \neq j} \mathbf{d_{ij}} \qquad (3)$$

where the sum of all possible paired-nodes distances is normalized over all the possible couples of N nodes.

Another useful metrics to quantify network robustness is the variation of the efficiency $\mathbf{\Delta E}$ depending on increasing number of removed nodes [179]. A robust network would have a small drop in the network efficiency with the removal of a node.

$$\mathbf{\Delta E_i} = \frac{\mathbf{E} - \mathbf{E_i}}{\mathbf{E}} \qquad (4)$$

where the efficiency is defined as $\mathbf{E} = \left(\mathbf{N} \cdot (\mathbf{N - 1})\right)^{-1} \sum_{i \neq j} \mathbf{d_{ij}^{-1}}$.

In the investigation presented here, we employ the theory of network robustness to assess the capability of networks to deliver and maintain an acceptable level of service in the presence of faults, as outlined by our models. Our analysis involves subjecting each network to both random and targeted attacks, following specific strategies (degree-, closeness-, and betweenness-based).

*Structural controllability and minimum driver nodes identification*

Lastly, we employ the structural controllability theory to assess the nodes ranking using the Kalman's rank condition for continuous linear time-invariant systems [157]. In addition, we also implement the Minimum Driver Nodes (MDN) algorithm [160,161], proposed by Liu et al., which is based on the minimal set of input signals required to fully control the network, and the MDN selection algorithm used by Zhang et al. [158], which can be used to identify the driver nodes — the nodes on which an input signal must be injected to obtain full control of the network.

Generally, the time-evolution of a network system consisting of N nodes and M input signals, with M ≤ N, can be described with the following linear differential equation

$$\frac{d\mathbf{x(t)}}{dt} = \mathbf{Ax(t)} + \mathbf{Bu(t)} \qquad (2)$$

where $\mathbf{x} = (\mathbf{x_1, x_2, x_3, \ldots, x_N})^T$ is the state vector for the N nodes system and $\mathbf{u} = (\mathbf{u_1, u_2, u_3, \ldots, u_M})^T$ is the control vector. A is the N×N state matrix, in which each element $\mathbf{a_{ij}}$ identifies the connection between the i[th] and j[th] nodes, while B is the M×N control matrix, whose dimension M depends on the number of input signals

$$\mathbf{B} = (\mathbf{e_1^T\ e_2^T\ e_3^T\ \ldots e_M^T}) \qquad (3)$$

where $\{\mathbf{e_1, e_2, e_3, \ldots, e_M}\}$ are the vectors of the canonical base.

Given A and B, it is possible to assemble the controllability matrix C:

$$\mathbf{C} = (\mathbf{B,\ AB,\ A^2B,\ A^3B,\ \ldots,\ A^{N-1}B}) \qquad (4)$$

If the controllability matrix has a full rank, i.e., rank(C) = N, the network is fully controllable. Theory and algorithms are well-explained in Ref. [97].

## References

1.  Schröder M, Kaufman RJ. THE MAMMALIAN UNFOLDED PROTEIN RESPONSE. Annu Rev Biochem. 2005 Jun 1;74(1):739–89.
2.  Malhotra JD, Kaufman RJ. The endoplasmic reticulum and the unfolded protein response. Semin Cell Dev Biol. 2007 Dec;18(6):716–31.
3.  Määttänen P, Gehring K, Bergeron JJM, Thomas DY. Protein quality control in the ER: The recognition of misfolded proteins. Semin Cell Dev Biol. 2010 Jul;21(5):500–11.
4.  Stolz A, Wolf DH. Endoplasmic reticulum associated protein degradation: A chaperone assisted journey to hell. Biochimica et Biophysica Acta (BBA) - Molecular Cell Research. 2010 Jun;1803(6):694–705.
5.  Uversky VN. Natively unfolded proteins: A point where biology waits for physics. Protein Science. 2002 Apr;11(4):739–56.
6.  Hetz C, Papa FR. The Unfolded Protein Response and Cell Fate Control. Mol Cell. 2018 Jan;69(2):169–81.
7.  Hetz C. The unfolded protein response: controlling cell fate decisions under ER stress and beyond. Nat Rev Mol Cell Biol. 2012 Feb;13(2):89–102.
8.  Wang J, Lee J, Liem D, Ping P. HSPA5 Gene encoding Hsp70 chaperone BiP in the endoplasmic reticulum. Gene. 2017 Jun;618:14–23.
9.  Kopp MC, Larburu N, Durairaj V, Adams CJ, Ali MMU. UPR proteins IRE1 and PERK switch BiP from chaperone to ER stress sensor. Nat Struct Mol Biol. 2019 Nov;26(11):1053–62.
10.  Shen J, Chen X, Hendershot L, Prywes R. ER Stress Regulation of ATF6 Localization by Dissociation of BiP/GRP78 Binding and Unmasking of Golgi Localization Signals. Dev Cell. 2002 Jul;3(1):99–111.
11.  Shen J, Snapp EL, Lippincott-Schwartz J, Prywes R. Stable Binding of ATF6 to BiP in the Endoplasmic Reticulum Stress Response. Mol Cell Biol. 2005 Feb;25(3):921–32.
12.  Sharma P, Alizadeh J, Juarez M, Samali A, Halayko AJ, Kenyon NJ, et al. Autophagy, Apoptosis, the Unfolded Protein Response, and Lung Function in Idiopathic Pulmonary Fibrosis. Cells. 2021 Jun;10(7):1642.
13.  Brown MK, Naidoo N. The endoplasmic reticulum stress response in aging and age-related diseases. Front Physiol. 2012;3.
14.  Brown MK, Chan MT, Zimmerman JE, Pack AI, Jackson NE, Naidoo N. Aging induced endoplasmic

reticulum stress alters sleep and sleep homeostasis. Neurobiol Aging. 2014 Jun;35(6):1431–41.

15.     Bommiasamy H, Popko B. Animal Models in the Study of the Unfolded Protein Response. In 2011. p. 91–109.

16.     Hollien J. Evolution of the unfolded protein response. Biochimica et Biophysica Acta (BBA) - Molecular Cell Research. 2013 Nov;1833(11):2458–63.

17.     Mori K. Evolutionary Aspects of the Unfolded Protein Response. Cold Spring Harb Perspect Biol. 2022 Aug 8;a041262.

18.     Krishnan K, Askew DS. The fungal UPR. Virulence. 2014 Feb 15;5(2):334–40.

19.     Moreno AA, Orellana A. The physiological role of the unfolded protein response in plants. Biol Res. 2011;44(1):75–80.

20.     Samperna S, Boari A, Vurro M, Salzano AM, Reveglia P, Evidente A, et al. Arabidopsis Defense against the Pathogenic Fungus Drechslera gigantea Is Dependent on the Integrity of the Unfolded Protein Response. Biomolecules. 2021 Feb 8;11(2):240.

21.     Raman K. Construction and analysis of protein–protein interaction networks. Autom Exp. 2010;2(1):2.

22.     Tomkins JE, Manzoni C. Advances in protein-protein interaction network analysis for Parkinson's disease. Neurobiol Dis. 2021 Jul;155:105395.

23.     Safari-Alighiarloo N, Taghizadeh M, Rezaei-Tavirani M, Goliaei B, Peyvandi AA. Protein-protein interaction networks (PPI) and complex diseases. Gastroenterol Hepatol Bed Bench. 2014;7(1):17–31.

24.     Barabási AL, Oltvai ZN. Network biology: understanding the cell's functional organization. Nat Rev Genet. 2004 Feb;5(2):101–13.

25.     Scheper W, Hoozemans JJM. The unfolded protein response in neurodegenerative diseases: a neuropathological perspective. Acta Neuropathol. 2015 Sep;130(3):315–31.

26.     van Ziel AM, Scheper W. The UPR in Neurodegenerative Disease: Not Just an Inside Job. Biomolecules. 2020 Jul;10(8):1090.

27.     Ghemrawi R, Khair M. Endoplasmic Reticulum Stress and Unfolded Protein Response in Neurodegenerative Diseases. Int J Mol Sci. 2020 Aug;21(17):6127.

28.     Glover K, Silverman L. Characterization of structural controllability. IEEE Trans Automat Contr. 1976 Aug;21(4):534–7.

29.     Shields R, Pearson J. Structural controllability of multiinput linear systems. IEEE Trans Automat Contr. 1976 Apr;21(2):203–12.

30.     Ching-Tai Lin. Structural controllability. IEEE Trans Automat Contr. 1974 Jun;19(3):201–8.

31.     Uhart M, Flores G, Bustos DM. Controllability of protein-protein interaction phosphorylation-based networks: Participation of the hub 14-3-3 protein family. Sci Rep. 2016 May;6(1):26234.

32.     Abdallah CT. Mathematical controllability of genomic networks. Proceedings of the National Academy of Sciences. 2011 Oct;108(42):17243–4.

33.     Vinayagam A, Gibson TE, Lee HJ, Yilmazel B, Roesel C, Hu Y, et al. Controllability analysis of the directed human protein interaction network identifies disease genes and drug targets. Proceedings of the National Academy of Sciences. 2016 May;113(18):4976–81.

34.     Liu S, Xu Q, Chen A, Wang P. Structural controllability of dynamic transcriptional regulatory networks for Saccharomyces cerevisiae. Physica A: Statistical Mechanics and its Applications. 2020 Jan;537:122772.

35.     Ackerman EE, Alcorn JF, Hase T, Shoemaker JE. A dual controllability analysis of influenza virus-host protein-protein interaction networks for antiviral drug target discovery. BMC Bioinformatics. 2019 Dec;20(1):297.

36.     Kanhaiya K, Czeizler E, Gratie C, Petre I. Controlling Directed Protein Interaction Networks in Cancer. Sci Rep. 2017 Sep;7(1):10327.

37.     Gonzalez O. Protein–Protein Interaction Databases. In: Encyclopedia of Systems Biology. New York, NY: Springer New York; 2013. p. 1786–90.

38.     Bajpai AK, Davuluri S, Tiwary K, Narayanan S, Oguru S, Basavaraju K, et al. Systematic comparison of the protein-protein interaction databases from a user's perspective. J Biomed Inform. 2020 Mar;103:103380.

39.     Nakajima N, Akutsu T, Nakato R. Databases for Protein–Protein Interactions. In 2021. p. 229–48.

40.     Lehne B, Schlitt T. Protein-protein interaction databases: keeping up with growing interactomes. Hum Genomics. 2009 Dec;3(3):291.

41.     Liu YY, Barabási AL. Control principles of complex systems. Rev Mod Phys. 2016 Sep;88(3):035006.

42.     Liu YY, Slotine JJ, Barabási AL. Controllability of complex networks. Nature. 2011 May;473(7346):167–73.

43.     Liu J, Zhou M, Wang S, Liu P. A comparative study of network robustness measures. Front Comput Sci. 2017 Aug;11(4):568–84.

44.     Oehlers M, Fabian B. Graph Metrics for Network Robustness—A Survey. Mathematics. 2021 Apr 17;9(8):895.

45. Artime O, Grassia M, De Domenico M, Gleeson JP, Makse HA, Mangioni G, et al. Robustness and resilience of complex networks. Nature Reviews Physics. 2024 Jan 8;

46. Barthelemy M. Betweenness Centrality. In 2018. p. 51–73.

47. Smith KM, Escudero J. Normalised degree variance. Appl Netw Sci. 2020 Dec 22;5(1):32.

48. Zhang P, Tao L, Zeng X, Qin C, Chen S, Zhu F, et al. A protein network descriptor server and its use in studying protein, disease, metabolic and drug targeted networks. Brief Bioinform. 2016 Aug 19;bbw071.

49. Bookstein A, Kulyukin VA, Raita T. Generalized Hamming Distance. Inf Retr Boston. 2002;5(4):353–75.

50. Vowels JJ, Thomas JH. Multiple chemosensory defects in daf-11 and daf-21 mutants of Caenorhabditis elegans. Genetics. 1994 Oct 1;138(2):303–16.

51. Birnby DA, Link EM, Vowels JJ, Tian H, Colacurcio PL, Thomas JH. A Transmembrane Guanylyl Cyclase (DAF-11) and Hsp90 (DAF-21) Regulate a Common Set of Chemosensory Behaviors in *Caenorhabditis elegans*. Genetics. 2000 May 1;155(1):85–104.

52. Hetz C, Zhang K, Kaufman RJ. Mechanisms, regulation and functions of the unfolded protein response. Nat Rev Mol Cell Biol. 2020 Aug;21(8):421–38.

53. Du Z, Chakrabarti S, Kulaberoglu Y, Smith ESJ, Dobson CM, Itzhaki LS, et al. Probing the unfolded protein response in long-lived naked mole-rats. Biochem Biophys Res Commun. 2020 Sep;529(4):1151–7.

54. Timberlake M, Prall K, Roy B, Dwivedi Y. Unfolded protein response and associated alterations in toll-like receptor expression and interaction in the hippocampus of restraint rats. Psychoneuroendocrinology. 2018 Mar;89:185–93.

55. Naidoo N. The Unfolded Protein Response in Mouse Cerebral Cortex. In 2011. p. 3–21.

56. Rana T, Shinde VM, Starr CR, Kruglov AA, Boitet ER, Kotla P, et al. An activated unfolded protein response promotes retinal degeneration and triggers an inflammatory response in the mouse retina. Cell Death Dis. 2014 Dec;5(12):e1578–e1578.

57. Kubra KT, Uddin MA, Akhter MS, Barabutis N. Hsp90 inhibitors induce the unfolded protein response in bovine and mice lung cells. Cell Signal. 2020 Mar;67:109500.

58. Umeda S, Suzuki MT, Okamoto H, Ono F, Mizota A, Terao K, et al. Molecular composition of drusen and possible involvement of anti-retinal autoimmunity in two different forms of macular degeneration in cynomolgus monkey ( *Macaca fascicularis* ). The FASEB Journal. 2005 Oct 12;19(12):1683–5.

59. Yonekura S, Tsuchiya M, Tokutake Y, Mizusawa M, Nakano M, Miyaji M, et al. The unfolded protein response is involved in both differentiation and apoptosis of bovine mammary epithelial cells. J Dairy Sci. 2018 Apr;101(4):3568–78.

60. Ghribi O, Herman MM, Pramoonjago P, Savory J. MPP $^+$ Induces the Endoplasmic Reticulum Stress Response in Rabbit Brain Involving Activation of the ATF-6 and NF-κB Signaling Pathways. J Neuropathol Exp Neurol. 2003 Nov 1;62(11):1144–53.

61. Kruzliak P, Sabo J, Zulli A. Endothelial endoplasmic reticulum and nitrative stress in endothelial dysfunction in the atherogenic rabbit model. Acta Histochem. 2015 Oct;117(8):762–6.

62. Huo Y, Ma F, Li T, Lei C, Liao J, Han Q, et al. Exposure to copper activates mitophagy and endoplasmic reticulum stress-mediated apoptosis in chicken ( *Gallus gallus* ) cerebrum. Environ Toxicol. 2023 Feb 9;38(2):392–402.

63. Gao PC, Wang AQ, Chen XW, Cui H, Li Y, Fan RF. Selenium alleviates endoplasmic reticulum calcium depletion-induced endoplasmic reticulum stress and apoptosis in chicken myocardium after mercuric chloride exposure. Environmental Science and Pollution Research. 2023 Feb 22;30(18):51531–41.

64. Lin YF, Sam J, Evans T. Sirt1 promotes tissue regeneration in zebrafish through regulating the mitochondrial unfolded protein response. iScience. 2021 Oct;24(10):103118.

65. Vacaru AM, Di Narzo AF, Howarth DL, Tsedensodnom O, Imrie D, Cinaroglu A, et al. Molecularly defined unfolded protein response subclasses have distinct correlations with fatty liver disease in zebrafish. Dis Model Mech. 2014 Jul 1;7(7):823–35.

66. Li J, Chen Z, Gao LY, Colorni A, Ucko M, Fang S, et al. A transgenic zebrafish model for monitoring xbp1 splicing and endoplasmic reticulum stress in vivo. Mech Dev. 2015 Aug;137:33–44.

67. Ryoo HD. Drosophila as a model for unfolded protein response research. BMB Rep. 2015 Aug;48(8):445–53.

68. Katow H, Vasudevan D, Ryoo HD. Drosophila Unfolded Protein Response (UPR) Assays In Vitro and In Vivo. In 2022. p. 261–77.

69. Demay Y, Perochon J, Szuplewski S, Mignotte B, Gaumer S. The PERK pathway independently triggers apoptosis and a Rac1/Slpr/JNK/Dilp8 signaling favoring tissue homeostasis in a chronic ER stress Drosophila model. Cell Death Dis. 2014 Oct;5(10):e1452–e1452.

70. Shen X, Ellis RE, Lee K, Liu CY, Yang K, Solomon A, et al. Complementary Signaling Pathways Regulate the Unfolded Protein Response and Are Required for C. elegans Development. Cell. 2001

Dec;107(7):893–903.

71.  Beaudoin-Chabot C, Wang L, Celik C, Abdul Khalid ATF, Thalappilly S, Xu S, et al. The unfolded protein response reverses the effects of glucose on lifespan in chemically-sterilized C. elegans. Nat Commun. 2022 Oct;13(1):5889.

72.  Urano F, Calfon M, Yoneda T, Yun C, Kiraly M, Clark SG, et al. A survival pathway for *Caenorhabditis elegans* with a blocked unfolded protein response. J Cell Biol. 2002 Aug;158(4):639–46.

73.  Kimata Y, Ishiwata-Kimata Y, Yamada S, Kohno K. Yeast unfolded protein response pathway regulates expression of genes for anti-oxidative stress and for cell surface proteins. Genes to Cells. 2006 Jan;11(1):59–69.

74.  Nguyen PTM, Ishiwata-Kimata Y, Kimata Y. Fast-Growing Saccharomyces cerevisiae Cells with a Constitutive Unfolded Protein Response and Their Potential for Lipidic Molecule Production. Appl Environ Microbiol. 2022 Nov;88(21).

75.  Ruberti C, Brandizzi F. Unfolded Protein Response in Arabidopsis. Methods Mol Biol. 2018;1691:231–8.

76.  Manghwar H, Li J. Endoplasmic Reticulum Stress and Unfolded Protein Response Signaling in Plants. Int J Mol Sci. 2022 Jan 13;23(2).

77.  Kamauchi S, Nakatani H, Nakano C, Urade R. Gene expression in response to endoplasmic reticulum stress in *Arabidopsis thaliana*. FEBS J. 2005 Jul 24;272(13):3461–76.

78.  Alcântara A, Seitner D, Navarrete F, Djamei A. A high-throughput screening method to identify proteins involved in unfolded protein response of the endoplasmic reticulum in plants. Plant Methods. 2020 Dec 21;16(1):4.

79.  Rehman S ur, Nadeem A, Javed M, Hassan F ul, Luo X, Khalid RB, et al. Genomic Identification, Evolution and Sequence Analysis of the Heat-Shock Protein Gene Family in Buffalo. Genes (Basel). 2020 Nov;11(11):1388.

80.  Turan M. Genome-wide analysis and characterization of HSP gene families (HSP20, HSP40, HSP60, HSP70, HSP90) in the yellow fever mosquito *(Aedes aegypti)* (Diptera: Culicidae). Journal of Insect Science. 2023 Nov;23(6).

81.  Storey JM, Storey KB. Chaperone proteins: universal roles in surviving environmental stress. Cell Stress Chaperones. 2023 Sep;28(5):455–66.

82.  Whitley D, Goldberg SP, Jordan WD. Heat shock proteins: A review of the molecular chaperones. J Vasc Surg. 1999 Apr;29(4):748–51.

83.  Hu C, Yang J, Qi Z, Wu H, Wang B, Zou F, et al. Heat shock proteins: Biological functions, pathological roles, and therapeutic opportunities. MedComm (Beijing). 2022 Sep;3(3).

84.  Hu C, Yang J, Qi Z, Wu H, Wang B, Zou F, et al. Heat shock proteins: Biological functions, pathological roles, and therapeutic opportunities. MedComm (Beijing). 2022 Sep 2;3(3).

85.  Sun H, Zou HY, Cai XY, Zhou HF, Li XQ, Xie WJ, et al. Network Analyses of the Differential Expression of Heat Shock Proteins in Glioma. DNA Cell Biol. 2020 Jul 1;39(7):1228–42.

86.  Liu H, Xiao F, Serebriiskii IG, O'Brien SW, Maglaty MA, Astsaturov I, et al. Network Analysis Identifies an HSP90-Central Hub Susceptible in Ovarian Cancer. Clinical Cancer Research. 2013 Sep 15;19(18):5053–67.

87.  Manikandan P, Vijayakumar R, Alshehri B, Senthilkumar S, Al-Aboody MS, Veluchamy A, et al. Exploring the biological behavior of Heat shock protein (HSPs) for understanding the Anti-ischemic stroke in humans. J Infect Public Health. 2022 Apr;15(4):379–88.

88.  Sun H, Cai X, Zhou H, Li X, Du Z, Zou H, et al. The protein–protein interaction network and clinical significance of heat-shock proteins in esophageal squamous cell carcinoma. Amino Acids. 2018 Jun 27;50(6):685–97.

89.  Park SM, Kang TI, So JS. Roles of XBP1s in Transcriptional Regulation of Target Genes. Biomedicines. 2021 Jul;9(7):791.

90.  Wang FM, Chen YJ, Ouyang HJ. Regulation of unfolded protein response modulator XBP1s by acetylation and deacetylation. Biochemical Journal. 2011 Jan;433(1):245–52.

91.  Hillary RF, FitzGerald U. A lifetime of stress: ATF6 in development and homeostasis. J Biomed Sci. 2018 Dec;25(1):48.

92.  Tam AB, Roberts LS, Chandra V, Rivera IG, Nomura DK, Forbes DJ, et al. The UPR Activator ATF6 Responds to Proteotoxic and Lipotoxic Stress by Distinct Mechanisms. Dev Cell. 2018 Aug;46(3):327-343.e7.

93.  Gardner BM, Pincus D, Gotthardt K, Gallagher CM, Walter P. Endoplasmic Reticulum Stress Sensing in the Unfolded Protein Response. Cold Spring Harb Perspect Biol. 2013 Mar;5(3):a013169–a013169.

94.  Nowakowska M, Gualtieri F, von Rüden EL, Hansmann F, Baumgärtner W, Tipold A, et al. Profiling the Expression of Endoplasmic Reticulum Stress Associated Heat Shock Proteins in Animal Epilepsy Models. Neuroscience. 2020 Mar;429:156–72.

95.    Chen Y, Mi Y, Zhang X, Ma Q, Song Y, Zhang L, et al. Dihydroartemisinin-induced unfolded protein response feedback attenuates ferroptosis via PERK/ATF4/HSPA5 pathway in glioma cells. Journal of Experimental & Clinical Cancer Research. 2019 Dec;38(1):402.

96.    Chen X, Shi C, He M, Xiong S, Xia X. Endoplasmic reticulum stress: molecular mechanism and therapeutic targets. Signal Transduct Target Ther. 2023 Sep;8(1):352.

97.    Luchetti N, Loppini A, Matarrese MAG, Chiodo L, Filippi S. Structural controllability to unveil hidden regulation mechanisms in Unfolded Protein Response: The role of network models. Physica A: Statistical Mechanics and its Applications. 2023 May;617:128671.

98.    Cox JS, Shamu CE, Walter P. Transcriptional induction of genes encoding endoplasmic reticulum resident proteins requires a transmembrane protein kinase. Cell. 1993 Jun;73(6):1197–206.

99.    Morl K, Ma W, Gething MJ, Sambrook J. A transmembrane protein with a cdc2+CDC28-related kinase activity is required for signaling from the ER to the nucleus. Cell. 1993 Aug;74(4):743–56.

100.    Riaz TA, Junjappa RP, Handigund M, Ferdous J, Kim HR, Chae HJ. Role of Endoplasmic Reticulum Stress Sensor IRE1α in Cellular Physiology, Calcium, ROS Signaling, and Metaflammation. Cells. 2020 May 8;9(5):1160.

101.    Trentmann SM. ERN1, a novel ethylene-regulated nuclear protein of Arabidopsis. Plant Mol Biol. 2000;44(1):11–25.

102.    Angelos E, Brandizzi F. The UPR regulator IRE1 promotes balanced organ development by restricting TOR-dependent control of cellular differentiation in Arabidopsis. The Plant Journal. 2022 Mar 18;109(5):1229–48.

103.    Ruberti C, Kim SJ, Stefano G, Brandizzi F. Unfolded protein response in plants: one master, many questions. Curr Opin Plant Biol. 2015 Oct;27:59–66.

104.    Duan G, Wu G, Chen X, Tian D, Li Z, Sun Y, et al. HGD: an integrated homologous gene database across multiple species. Nucleic Acids Res. 2023 Jan 6;51(D1):D994–1002.

105.    Bateman A, Martin MJ, Orchard S, Magrane M, Ahmad S, Alpi E, et al. UniProt: the Universal Protein Knowledgebase in 2023. Nucleic Acids Res. 2023 Jan;51(D1):D523–31.

106.    The UniProt Consortium. UniProt: a worldwide hub of protein knowledge. Nucleic Acids Res. 2019 Jan;47(D1):D506–15.

107.    Wang Y, Wang Q, Huang H, Huang W, Chen Y, McGarvey PB, et al. A crowdsourcing open platform for literature curation in UniProt. PLoS Biol. 2021 Dec;19(12):e3001464.

108.    Cohen R, Erez K, ben-Avraham D, Havlin S. Resilience of the Internet to Random Breakdowns. Phys Rev Lett. 2000 Nov 20;85(21):4626–8.

109.    Cohen R, Erez K, ben-Avraham D, Havlin S. Breakdown of the Internet under Intentional Attack. Phys Rev Lett. 2001 Apr 16;86(16):3682–5.

110.    Holme P, Kim BJ, Yoon CN, Han SK. Attack vulnerability of complex networks. Phys Rev E. 2002 May 7;65(5):056109.

111.    Ruj S, Pal A. Analyzing Cascading Failures in Smart Grids under Random and Targeted Attacks. In: 2014 IEEE 28th International Conference on Advanced Information Networking and Applications. IEEE; 2014. p. 226–33.

112.    Callaway DS, Newman MEJ, Strogatz SH, Watts DJ. Network Robustness and Fragility: Percolation on Random Graphs. Phys Rev Lett. 2000 Dec 18;85(25):5468–71.

113.    Jeronimo C, Forget D, Bouchard A, Li Q, Chua G, Poitras C, et al. Systematic Analysis of the Protein Interaction Network for the Human Transcription Machinery Reveals the Identity of the 7SK Capping Enzyme. Mol Cell. 2007 Jul;27(2):262–74.

114.    Yuxiong W, Faping L, Bin L, Yanghe Z, Yao L, Yunkuo L, et al. Regulatory mechanisms of the cAMP-responsive element binding protein 3 (CREB3) family in cancers. Biomedicine & Pharmacotherapy. 2023 Oct;166:115335.

115.    Bunney TD, Cole AR, Broncel M, Esposito D, Tate EW, Katan M. Crystal Structure of the Human, FIC-Domain Containing Protein HYPE and Implications for Its Functions. Structure. 2014 Dec;22(12):1831–43.

116.    Sanyal A, Chen AJ, Nakayasu ES, Lazar CS, Zbornik EA, Worby CA, et al. A Novel Link between Fic (Filamentation Induced by cAMP)-mediated Adenylylation/AMPylation and the Unfolded Protein Response. Journal of Biological Chemistry. 2015 Mar;290(13):8482–99.

117.    Sato T, Sako Y, Sho M, Momohara M, Suico MA, Shuto T, et al. STT3B-Dependent Posttranslational N-Glycosylation as a Surveillance System for Secretory Protein. Mol Cell. 2012 Jul;47(1):99–110.

118.    Terrab L, Wipf P. Hsp70 and the Unfolded Protein Response as a Challenging Drug Target and an Inspiration for Probe Molecule Development. ACS Med Chem Lett. 2020 Mar 12;11(3):232–6.

119.    Heldens L, Hensen SMM, Onnekink C, van Genesen ST, Dirks RP, Lubsen NH. An atypical unfolded protein response in heat shocked cells. PLoS One. 2011;6(8):e23512.

120.    Leng X, Wang X, Pang W, Zhan R, Zhang Z, Wang L, et al. Evidence of a role for both anti-Hsp70 antibody and endothelial surface membrane Hsp70 in atherosclerosis. Cell Stress Chaperones. 2013

Jul;18(4):483–93.

121.  Liang C, Li L, Zhao H, Lan M, Tang Y, Zhang M, et al. Identification and expression analysis of heat shock protein family genes of gall fly (Procecidochares utilis) under temperature stress. Cell Stress Chaperones. 2023 May;28(3):303–20.

122.  Imai J, Yashiroda H, Maruya M, Yahara I, Tanaka K. Proteasomes and molecular chaperones: cellular machinery responsible for folding and destruction of unfolded proteins. Cell Cycle. 2003;2(6):585–90.

123.  Runtuwene LR, Kawashima S, Pijoh VD, Tuda JSB, Hayashida K, Yamagishi J, et al. The Lethal(2)-Essential-for-Life [L(2)EFL] Gene Family Modulates Dengue Virus Infection in Aedes aegypti. Int J Mol Sci. 2020 Oct 12;21(20):7520.

124.  Landis G, Shen J, Tower J. Gene expression changes in response to aging compared to heat stress, oxidative stress and ionizing radiation in Drosophila melanogaster. Aging. 2012 Nov 30;4(11):768–89.

125.  Tower J. Heat shock proteins and Drosophila aging. Exp Gerontol. 2011 May;46(5):355–62.

126.  Landis G, Shen J, Tower J. Gene expression changes in response to aging compared to heat stress, oxidative stress and ionizing radiation in Drosophila melanogaster. Aging. 2012 Nov 30;4(11):768–89.

127.  Morrow G, Tanguay RM. Drosophila melanogaster Hsp22: a mitochondrial small heat shock protein influencing the aging process. Front Genet. 2015 Mar 16;6.

128.  Urano F, Calfon M, Yoneda T, Yun C, Kiraly M, Clark SG, et al. A survival pathway for *Caenorhabditis elegans* with a blocked unfolded protein response. J Cell Biol. 2002 Aug 19;158(4):639–46.

129.  Sun J, Singh V, Kajino-Sakamoto R, Aballay A. Neuronal GPCR Controls Innate Immunity by Regulating Noncanonical Unfolded Protein Response Genes. Science (1979). 2011 May 6;332(6030):729–32.

130.  Sun J, Liu Y, Aballay A. Organismal regulation of XBP-1-mediated unfolded protein response during development and immune activation. EMBO Rep. 2012 Sep 13;13(9):855–60.

131.  Alcântara A, Seitner D, Navarrete F, Djamei A. A high-throughput screening method to identify proteins involved in unfolded protein response of the endoplasmic reticulum in plants. Plant Methods. 2020 Dec;16(1):4.

132.  Tran HC, Van Aken O. Mitochondrial unfolded protein-related responses across kingdoms: similar problems, different regulators. Mitochondrion. 2020 Jul;53:166–77.

133.  Walter P, Ron D. The Unfolded Protein Response: From Stress Pathway to Homeostatic Regulation. Science (1979). 2011 Nov;334(6059):1081–6.

134.  Almanza A, Carlesso A, Chintha C, Creedican S, Doultsinos D, Leuzzi B, et al. Endoplasmic reticulum stress signalling – from basic mechanisms to clinical applications. FEBS J. 2019 Jan;286(2):241–78.

135.  Vidal RL, Hetz C. Crosstalk between the UPR and autophagy pathway contributes to handling cellular stress in neurodegenerative disease. Autophagy. 2012 Jun;8(6):970–2.

136.  MORI K. The unfolded protein response: the dawn of a new field. Proceedings of the Japan Academy, Series B. 2015;91(9):469–80.

137.  Ma Y, Hendershot LM. The Unfolding Tale of the Unfolded Protein Response. Cell. 2001 Dec;107(7):827–30.

138.  Hollien J. Evolution of the unfolded protein response. Biochimica et Biophysica Acta (BBA) - Molecular Cell Research. 2013 Nov;1833(11):2458–60.

139.  Zhang L, Zhang C, Wang A. Divergence and Conservation of the Major UPR Branch IRE1-bZIP Signaling Pathway across Eukaryotes. Sci Rep. 2016 Jun 3;6(1):27362.

140.  Gómora-García JC, Gerónimo-Olvera C, Pérez-Martínez X, Massieu L. IRE1α RIDD activity induced under ER stress drives neuronal death by the degradation of 14-3-3 θ mRNA in cortical neurons during glucose deprivation. Cell Death Discov. 2021 Jun 3;7(1):131.

141.  Maurel M, Chevet E, Tavernier J, Gerlo S. Getting RIDD of RNA: IRE1 in cell fate regulation. Trends Biochem Sci. 2014 May;39(5):245–54.

142.  Hollien J, Lin JH, Li H, Stevens N, Walter P, Weissman JS. Regulated Ire1-dependent decay of messenger RNAs in mammalian cells. Journal of Cell Biology. 2009 Aug 10;186(3):323–31.

143.  Zheng Z, Wang G, Li L, Tseng J, Sun F, Chen X, et al. Transcriptional signatures of unfolded protein response implicate the limitation of animal models in pathophysiological studies. Environ Dis. 2016;1(1):24.

144.  Breschi A, Gingeras TR, Guigó R. Comparative transcriptomics in human and mouse. Nat Rev Genet. 2017 Jul 8;18(7):425–40.

145.  Cheng Y, Ma Z, Kim BH, Wu W, Cayting P, Boyle AP, et al. Principles of regulatory information conservation between mouse and human. Nature. 2014 Nov 20;515(7527):371–5.

146.  Lin S, Lin Y, Nery JR, Urich MA, Breschi A, Davis CA, et al. Comparison of the transcriptional landscapes between human and mouse tissues. Proceedings of the National Academy of Sciences. 2014 Dec;111(48):17224–9.

147.  Zhu F, Nair RR, Fisher EMC, Cunningham TJ. Humanising the mouse genome piece by piece. Nat

Commun. 2019 Apr 23;10(1):1845.

148. Monaco G, van Dam S, Casal Novo Ribeiro JL, Larbi A, de Magalhães JP. A comparison of human and mouse gene co-expression networks reveals conservation and divergence at the tissue, pathway and disease levels. BMC Evol Biol. 2015 Dec 20;15(1):259.

149. Shih J, Hodge R, Andrade-Navarro MA. Comparison of inter- and intraspecies variation in humans and fruit flies. Genom Data. 2015 Mar;3:49–54.

150. Curcio R, Lunetti P, Zara V, Ferramosca A, Marra F, Fiermonte G, et al. Drosophila melanogaster Mitochondrial Carriers: Similarities and Differences with the Human Carriers. Int J Mol Sci. 2020 Aug 22;21(17):6052.

151. Shen X, Ellis RE, Lee K, Liu CY, Yang K, Solomon A, et al. Complementary Signaling Pathways Regulate the Unfolded Protein Response and Are Required for C. elegans Development. Cell. 2001 Dec;107(7):893–903.

152. Tran DM, Kimata Y. The unfolded protein response of yeast <i>Saccharomyces cerevisiae</i> and other organisms. PLANT MORPHOLOGY. 2018;30(1):15–24.

153. Klein B, Holmér L, Smith KM, Johnson MM, Swain A, Stolp L, et al. A computational exploration of resilience and evolvability of protein–protein interaction networks. Commun Biol. 2021 Dec;4(1):1352.

154. Ortiz-Vilchis P, De-la-Cruz-García JS, Ramirez-Arellano A. Identification of Relevant Protein Interactions with Partial Knowledge: A Complex Network and Deep Learning Approach. Biology (Basel). 2023 Jan;12(1):140.

155. Wagner A. Robustness and evolvability: a paradox resolved. Proceedings of the Royal Society B: Biological Sciences. 2008 Jan;275(1630):91–100.

156. Zitnik M, Sosič R, Feldman MW, Leskovec J. Evolution of resilience in protein interactomes across the tree of life. Proceedings of the National Academy of Sciences. 2019 Mar;116(10):4426–33.

157. Kalman RE. Mathematical Description of Linear Dynamical Systems. Journal of the Society for Industrial and Applied Mathematics Series A Control. 1963 Jan;1(2):152–92.

158. Zhang P, Ji Z, Li Z. Minimum driver nodes selection in complex networks. In: 2017 36th Chinese Control Conference (CCC). IEEE; 2017. p. 8461–6.

159. Aric A. Hagberg, Daniel A. Schult, Peter J. Swart. Exploring network structure, dynamics, and function using NetworkX. In: Gäel Varoquaux, Travis Vaught, Jarrod Millman, editors. Proceedings of the 7th Python in Science Conference (SciPy2008). Pasadena; 2008. p. 11–5.

160. Patel TP, Man K, Firestein BL, Meaney DF. Automated quantification of neuronal networks and single-cell calcium dynamics using calcium imaging. J Neurosci Methods. 2015 Mar;243:26–38.

161. Yuan Z, Zhao C, Di Z, Wang WX, Lai YC. Exact controllability of complex networks. Nat Commun. 2013 Sep;4(1):2447.

162. Chen C, Huang H, Wu CH. Protein Bioinformatics Databases and Resources. In 2017. p. 3–39.

163. Szklarczyk D, Kirsch R, Koutrouli M, Nastou K, Mehryary F, Hachilif R, et al. The STRING database in 2023: protein–protein association networks and functional enrichment analyses for any sequenced genome of interest. Nucleic Acids Res. 2023 Jan;51(D1):D638–46.

164. von Mering C. STRING: known and predicted protein-protein associations, integrated and transferred across organisms. Nucleic Acids Res. 2004 Dec;33(Database issue):D433–7.

165. Mering C v. STRING: a database of predicted functional associations between proteins. Nucleic Acids Res. 2003 Jan;31(1):258–61.

166. Snel B. STRING: a web-server to retrieve and display the repeatedly occurring neighbourhood of a gene. Nucleic Acids Res. 2000 Sep;28(18):3442–4.

167. Dasgupta A, Kumar R, Sarlos T. On estimating the average degree. In: Proceedings of the 23rd international conference on World wide web. New York, NY, USA: ACM; 2014. p. 795–806.

168. Chen HH, Dietrich U. Normalized closeness centrality of urban networks: impact of the location of the catchment area and evaluation based on an idealized network. Appl Netw Sci. 2023 Sep 6;8(1):60.

169. Mall R, Cerulo L, Bensmail H, Iavarone A, Ceccarelli M. Detection of statistically significant network changes in complex biological networks. BMC Syst Biol. 2017 Dec;11(1):32.

170. Van Rossum G, Drake FL. Python 3 Reference Manual. Scotts Valley, CA: CreateSpace; 2009.

171. The MathWorks Inc. MATLAB Version: 9.14.0.2254940 (R2023a) Update 2. Natick, Massachusetts, United States: The MathWorks Inc.; 2023.

172. Shaffer JP. Multiple Hypothesis Testing. Annu Rev Psychol. 1995 Jan;46(1):561–84.

173. Armstrong RA. When to use the <scp>B</scp> onferroni correction. Ophthalmic and Physiological Optics. 2014 Sep 2;34(5):502–8.

174. Lee S, Lee DK. What is the proper way to apply the multiple comparison test? Korean J Anesthesiol. 2018 Oct;71(5):353–60.

175. Curran-Everett D. Multiple comparisons: philosophies and illustrations. American Journal of Physiology-Regulatory, Integrative and Comparative Physiology. 2000 Jul 1;279(1):R1–8.

176. Ludbrook J. MULTIPLE COMPARISON PROCEDURES UPDATED. Clin Exp Pharmacol Physiol.

1998 Dec 28;25(12):1032–7.

177.     Bretz F, Posch M, Glimm E, Klinglmueller F, Maurer W, Rohmeyer K. Graphical approaches for multiple comparison procedures using weighted Bonferroni, Simes, or parametric tests. Biometrical Journal. 2011 Nov 12;53(6):894–913.

178.     Midway S, Robertson M, Flinn S, Kaller M. Comparing multiple comparisons: practical guidance for choosing the best multiple comparisons test. PeerJ. 2020;8:e10387.

179.     Liu J, Zhou M, Wang S, Liu P. A comparative study of network robustness measures. Front Comput Sci. 2017 Aug 26;11(4):568–84.

180.     Iyer S, Killingback T, Sundaram B, Wang Z. Attack Robustness and Centrality of Complex Networks. PLoS One. 2013 Apr 2;8(4):e59613.

181.     Kang K. Analysis of evolutionary process of fog computing system based on BA and ER network hybrid model. Evol Intell. 2020 Mar 3;13(1):33–8.

# Supplementary material of "A statistical mechanics investigation of Unfolded Protein Response across organisms"

Nicole Luchetti[1,2,*], Keith M. Smith[3], Margherita A. G. Matarrese[1], Alessandro Loppini[4], Simonetta Filippi[1,5,6,*], Letizia Chiodo[1]

[1] Department of Engineering, Università Campus Bio-Medico di Roma, Via Alvaro del Portillo 21, 00128 Rome, Italy
[2] Center for Life Nano- & Neuro-Science, Istituto Italiano di Tecnologia, Viale Regina Elena 291, 00161 Rome, Italy
[3] Department of Sciences, University of Strathclyde, 16 Richmond Street, Glasgow, G1 1XQ, Scotland, UK
[4] Department of Medicine and Surgery, Università Campus Bio-Medico di Roma, Via Alvaro del Portillo 21, 00128 Rome, Italy
[5] National Institute of Optics, National Research Council, Largo Enrico Fermi 6, 50125 Florence, Italy
[6] International Center for Relativistic Astrophysics Network, Piazza della Repubblica 10, 65122 Pescara, Italy

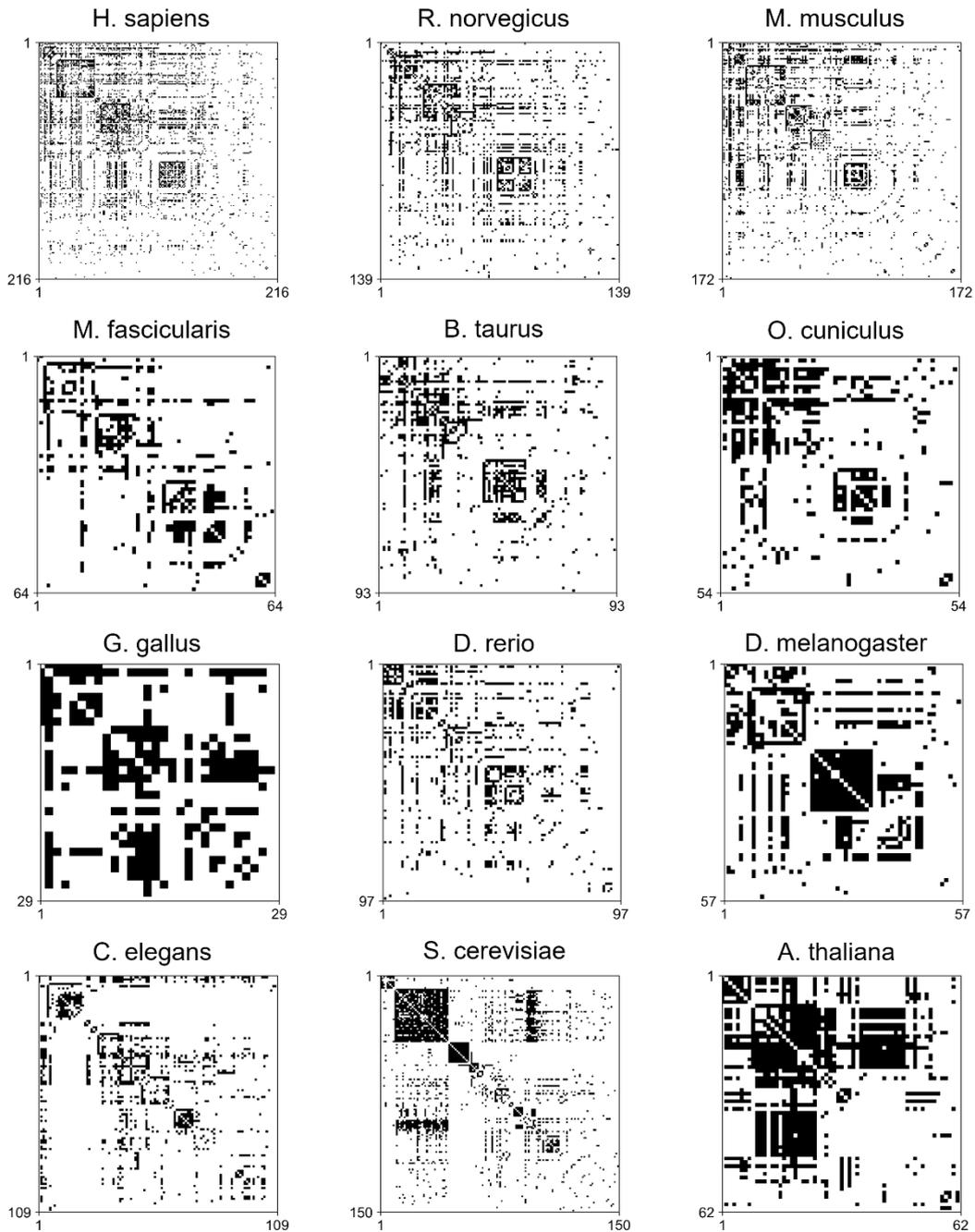

*Figure S1: Matrix representation of native models.*

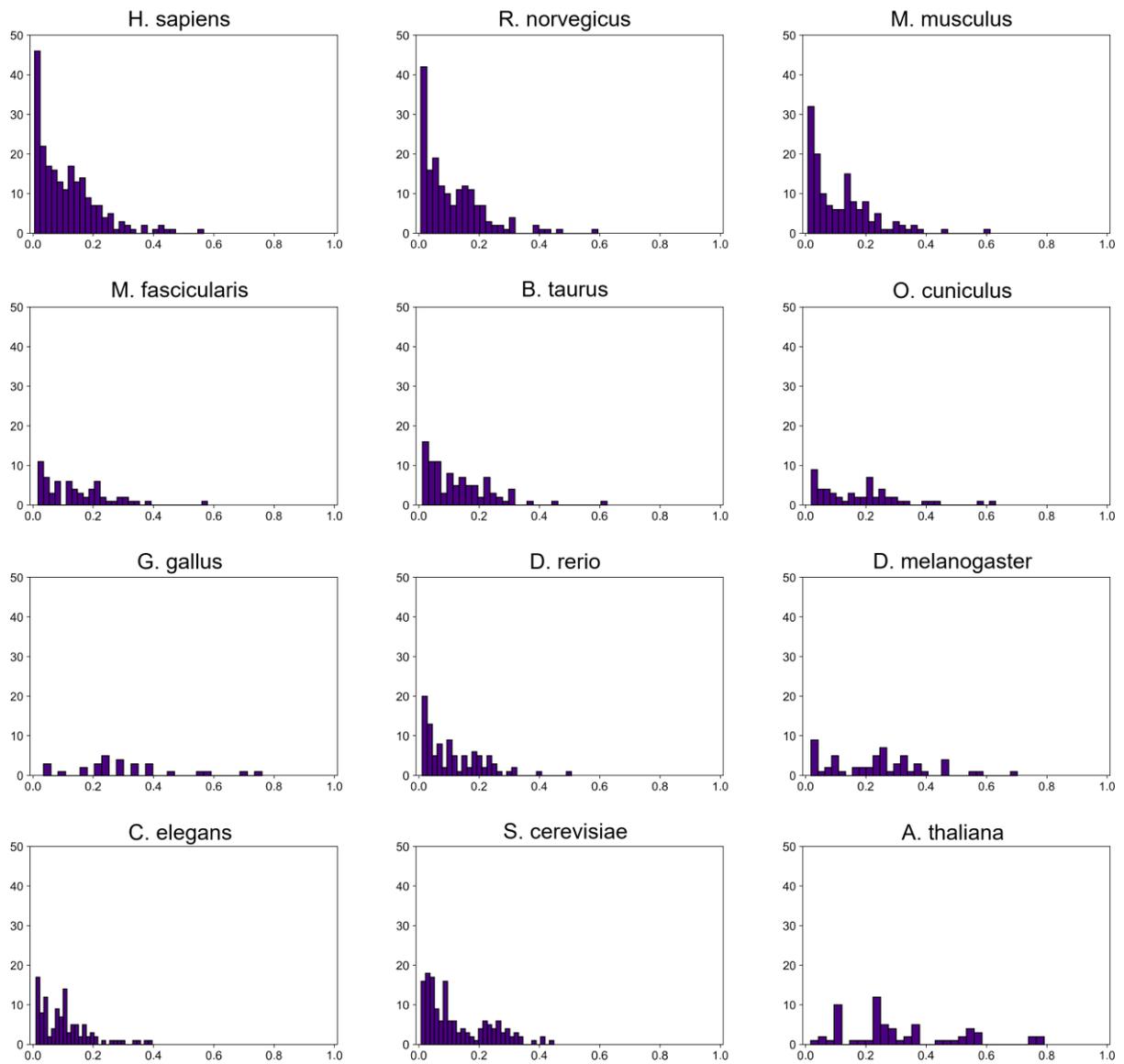

*Figure S2: Normalized node-degree distributions of native models.*

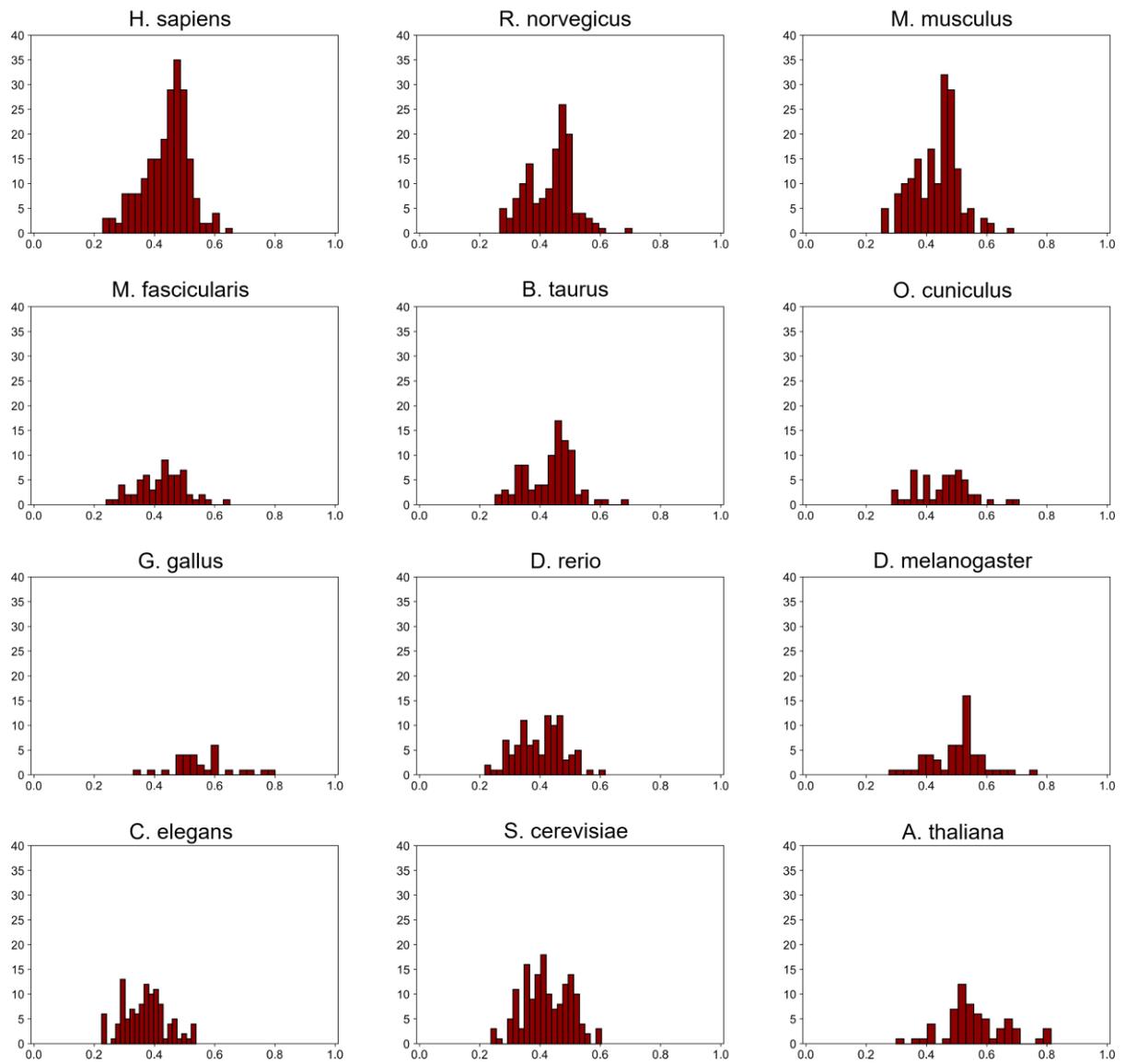

*Figure S3: Normalized closeness centrality distributions of native models.*

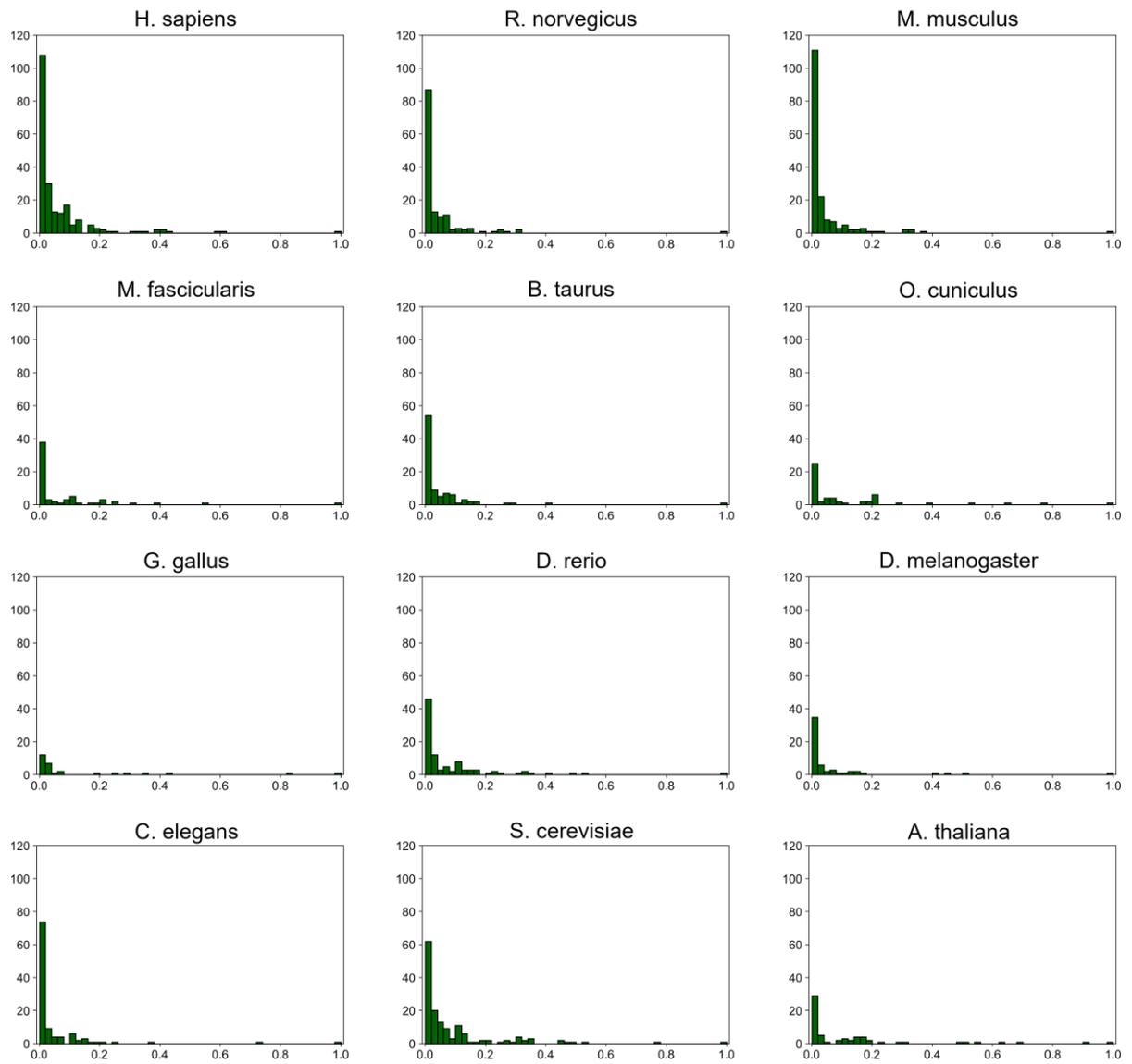

*Figure S4: Normalized betweenness centrality distributions of native models.*

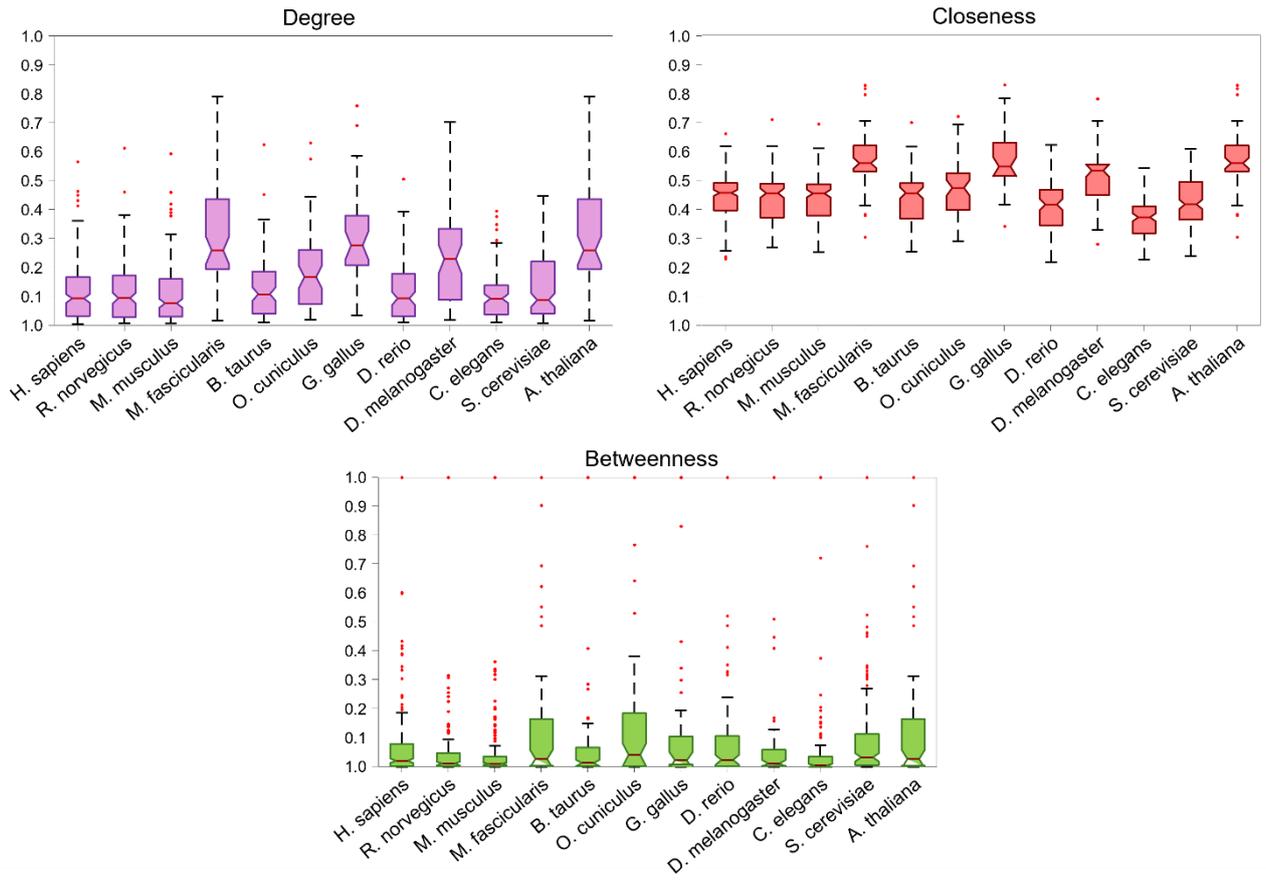

Figure S5: Representation of normalized network metrics.